\documentclass[letter,11pt]{article}
\usepackage[T1]{fontenc}

\usepackage{amsfonts,dif,amsmath}
\usepackage{latexsym}
\usepackage{color}
\usepackage{natbib}

\usepackage{verbatim}
\usepackage{graphicx}

\usepackage{float}

\renewcommand{\theequation}{\arabic{section}.\arabic{equation}}

\def\q{\quad}

\newtheorem{prop}{Proposition}[section]
\newtheorem{cor}[prop]{Corollary}

\newtheorem{lem}[prop]{Lemma}
\newtheorem{theo}[prop]{Theorem}
\newtheorem{rema}[prop]{Remark}

\newcommand{\etal}{{\em et al.}}
\graphicspath{{./images/}}

\renewcommand{\cite}{\citet*}

\begin{document}

\title{Statistical Properties of Microstructure Noise}

\author{Jean  Jacod,
Yingying Li
\footnote{Research partially supported by GRF 602710 of the HKSAR},
Xinghua Zheng
\footnote{Research partially supported by GRF 606811 of the HKSAR} }

\date{\today}

\maketitle

%\runauthor{J. Jacod, Y. Li and X. Zheng}

\begin{abstract}
We study the estimation of moments and joint moments of microstructure noise.
Estimators of arbitrary order of (joint) moments are provided, for which we
establish consistency as well as central limit theorems. In particular, we
provide estimators of auto-covariances and auto-correlations of the noise.
Simulation studies demonstrate excellent performance of our estimators even
in the presence of jumps and irregular observation times. Empirical studies
reveal (moderate) positive auto-correlation of the noise for the stocks
tested.

\end{abstract}


\section{Introduction}\label{sec-Intro}
\setcounter{equation}{0}
\renewcommand{\theequation}{\thesection.\arabic{equation}}

It has long been recognized that market microstructure noise plays a
significant role in financial markets. See, for example, the seminal paper
of \cite{Black} and comprehensive  reviews of \cite{Madhavan}, \cite{OHara},
\cite{Stoll} and \cite{Hasbrouck},  among others. The market microstructure
noise is induced by  various frictions in the trading process. Examples of
such frictions include bid-ask spread, asymmetric information of traders,
the discreteness of price change, etc.

With the increasing availability of high frequency data, the market
microstructure noise has received growing attention. Despite the small size,
market microstructure noise accumulates at high frequency and affects badly
the inferences about the efficient price processes, such as the estimation of
volatilities. No-arbitrage based arguments (see, for example,
\cite{delbaenschachermayer94}) suggests that the (efficient) price processes
should normally be semimartingales.  The fundamental properties of
semimartingales allow to make accurate inferences about volatilities and other
quantities with high frequency observations. See, for example,
\cite{jacodprotter98}, \cite{myklandzhang06}, among others. However, for
liquidly traded securities,   empirical evidence such as the signature plots
of \cite{abdl00} show clear noise accumulation effect at high frequency.
Therefore, more recent research carefully analyzes both components of the
market price processes: the  latent semimartingale  price process and the
noise process.

Several methods to de-noise the data in the context of volatility estimation
have been proposed. For example, the two-scale method as in
\cite{ZhangMA05}, \cite{AMZ05};  the kernel method as in \cite{BHLS08},
 \cite{BHLS11};  the pre-averaging method  as in  \cite{JLMPV-09},
 \cite{KinnebrockMC10}; the multi-scale method as in \cite{Zhang06};
  the quasi-maximum likelihood method as in \cite{Xiu10}, \cite{AFX},
   among others. These methods are shown to be very effective  when the
   noise is an additive white noise, or presents some kind of independence
    between successive observation times. \cite{gloterjacod01}, \cite{LiM07}
    and \cite{rosenbaum}
    studied the case when the noise is of specific form such as round-off
errors or round-off errors on top of additive white noise.
On the other hand, \cite{hansenlunde06} and \cite{UO09} have shown evidence
of dependence of the noise in financial markets.
When there is autocorrelation in the data, one possible way to
reduce the impact of dependence
is to use subsampling and averaging. \cite{hansenlunde06},
\cite{BHLS08} and \cite{yacperlan11} provide estimators when the noise satisfy
certain weak dependence assumptions.
However, the optimal subsampling scheme and de-noise method depend on the
dependence structure. Hence understanding the dependence structure of the
noise is essential for
inferences.

Can we understand better the statistical properties of the noise?
Specifically, for a particular security price process,  how is the noise
distributed and what is the dependence structure?

In this article we study how to estimate the moments and joint moments
of the noise,
based on high-frequency data. More specifically, under both settings where the
observation times are equally spaced or are
irregularly spaced, we propose estimators for (joint) moments of arbitrary
orders of the noise. We establish consistency
as well as central limit theorems for our estimators under certain mild mixing
conditions on the noise
(see Assumptions (NO-1) and (NO-2) below for precise statements). As is well
known that under {appropriate conditions on the tail} any distribution can
be fully reconstructed from its moments, our
results allow one to understand the
marginal distribution as well as the joint distributions of the noise.

Simple applications of our results include
estimating the auto-covariances and auto-correlations of the noise. And as
central limit theorems are available,
one can readily build tests for testing, for example, whether
auto-correlations of particular orders vanish or not.

The paper is organized as follows. Section \ref{sec:A} introduces the setting
and assumptions. Section \ref{sec:M} presents the consistency and asymptotic
normality results of our proposed estimators for the (joint) moments of the noise. Section \ref{sec:Sim} demonstrates
our results via simulations. Empirical studies are carried out in Section
\ref{sec:emp} in which we show by estimation and hypothesis testing that
the noises are (moderately) positively auto-correlated for the stocks tested. Section \ref{sec:C} concludes. The proofs are given in the Appendix \ref{sec:P}.

\section{Setting and assumptions}\label{sec:A}
\setcounter{equation}{0}
\renewcommand{\theequation}{\thesection.\arabic{equation}}

In this paper, we have three basic ingredients. The first one is the
underlying process $X$, typically the log-price of an asset; the
second one is the observation scheme, the third one is the noise.\\

The assumptions on $X$ are the standard ones in this kind of problem,
namely is  an It\^o semimartingale,  possibly discontinuous,   plus some mild
additional assumptions : so $X$ is defined on some
filtered probability space $\proba$, and it admits the following Grigelionis
representation:
\bee\label{S1}
X_t=X_0+\int_0^tb_sds+\int_0^t\si_sdW_s+(\de1_{\{|\de|\leq1\}})
\star(\um-\un)_t+(\de1_{\{|\de|>1\}})\star\um_t.
\eee
In this formula, $W$ is a standard Brownian motion, and $\um$ is a
Poisson random measure on $\R_+\times E$, where $(E,\ea)$ is a Polish
space, with a non-random intensity measure of the form
$\un(dt,dz)=dt\otimes\la(dz)$ with $\la$ a $\si$-finite measure on
$(E,\ea)$. The above is the general form of an It\^o semimartingale,
and we assume the following on the optional coefficients
$b$ and $\si$ and the predictable coefficient $\de=\de(\om,t,z)$:
\vsc

\nib Assumption (H): \rm The process $b$ is locally bounded,
the process $\si$ is c\`adl\`ag, and there is a localizing sequence
$(\tau_n)$ of stopping times and, for each $n$, a {\em deterministic}\,
nonnegative function $J_n$ on $E$ satisfying
$\int\! J_n(z)^2\,\la(dz)<\infty$
and such that $|\de(\om,t,z)|\wedge1\leq J_n(z)$ for all $(\om,t,z)$ with
$t\leq\tau_n(\om)$.
\vsc

Next, we describe how observations take place. At stage $n$, that is for
a given frequency of observations, the successive observations occur
at times $0=T(n,0)<T(n,1)<\cdots$, for a sequence $T(n,i)$ of (possibly
random) finite times increasing to $\infty$ as $i\to\infty$, so the number
of observations up to time $t$ is $N_n(t)+1$, where $N_n(t)=
\sup(i:T(n,i)\leq t)$. The minimal assumption on the observation times is
that each $T(n,i)$ is a stopping time, and the mesh goes to $0$ in a sense
specified later, as $n\to\infty$.
Moreover, at time $T(n,i)$ the process $X$ is
contaminated by some noise, meaning that we observe the variable
\bee\label{S3}
Y^n_i=X_{T(n,i)}+\ep^n_i,
\eee
where the noise is $\ep^n_i$.

For the sake of motivation about our forthcoming assumptions, we
(temporarily) suppose that the noise is independent of $X$, centered,
stationary, and with a negative exponential covariance. This covers a whole
range of ``natural'' situations, the two extreme ones being as follows:
\vsd

\ni 1) Conditionally on the observation times, the covariance
between $\ep^n_i$ and $\ep^n_{i+j}$ is $ae^{-a'(T(n,i+j)-T(n,i))}$: so
the exponential covariance is in terms of calendar time and does not depend
on the observation scheme.
\vsd

\ni 2) The  covariance between $\ep^n_i$ and $\ep^n_{i+j}$ is $ae^{-a'j}$:
so the covariance between two values of the noise depends only on
how many observations (or, transactions) occurred in between the
corresponding times.
\vsd

\ni And, of course, there are mid-term possibilities, like the covariance
being $ae^{-a'ju_n}$ with a ``scaling'' sequence   $u_n$ going to $0$
slower than $(T(n,i+1)-T(n,i))$.

In the first situation above it is of course impossible to obtain
consistent estimators for the characteristics of the noise, such as the
covariance function (in the negative exponential case, as well as in a
completely general case), unless the horizon $T$ goes to infinity. Such a
setting has  been studied, see e.g. \cite{UO09}, and here we are
interested in the case where the horizon $T$ is fixed. In the second
extremal situation, and in all intermediate cases, it is in principle
possible to consistently estimate the characteristics of the noise, under
appropriate assumptions of course. The extreme case 2 above is obviously
simpler than the intermediate cases, and should already provide useful
insight, so below we focus on the second extremal case.

Now, it is well known that, even in the absence of noise,   analysis of the
underlying process such as the estimation of
the volatility is much easier when observation times are equally spaced,
that is $T(n,i)=i\De_n$ for a sequence of non-random numbers $\De_n$ going
to $0$. And, in this case, when there is noise we can relax somehow the
independence assumption between $X$ and the noise. On the other hand, when
the noise is indeed independent of~$X$, the statistical analysis of the
noise does not require equally spaced observations: this is especially
interesting when observation times coincide with transaction times,
those being of course not equally spaced (and the extremal case described
above is then rather well suited to real problems).

This is why, below, we consider two different sets of assumptions,
which combine hypotheses on the observation scheme and on the structure
of the noise.

Before stating these assumptions, and for completeness, we recall
the $\rho$-mixing property of a stationary sequence $(\chi_i)_{i\in\Z}$
of variables, indexed by $\Z$: letting $\g_j=\si(\chi_i:\,i\leq j)$ and
$\g^j=\si(\chi_i:\,i\geq j)$ be the pre- and post-$\si$-fields at time
$j$, the $\rho$-mixing coefficients of $\chi$ for $k\geq1$ are
\bee\label{S0}
\begin{array}{ll}
\rho_k(\chi)=\sup\big(|\E(UV)|:&\E(U)=\E(V)=0,~\E(U^2)\leq1,~
\E(V^2)\leq1,\\
\hskip3cm~&\text{$U$ is $\g_0$-measurable, $V$ is $\g^k$-measurable}),
\end{array}
\eee
and we say that $\chi$ is $v$-polynomially $\rho$-mixing if
$\rho_k(\chi)\leq K/k^v$, where $v$ is a number bigger than $1$.
Then, the two sets of assumptions are as follows, and both of them make
use of a {\em non-random} sequence of positive numbers $\De_n$ going
to $0$ as $n\to\infty$.
\vsc

\nib Assumption (NO-1): \rm For all $T>0$ we have
\bee\label{S2}
\begin{array}{c}
\text{the sequences}~~
\frac1{\De_n}\,\sup\nolimits_{i\geq1}\,\big(T(n,i)\wedge T
-T(n,i-1)\wedge T\big)\\
~~\text{and}~~\De_nN_n(T)
\text{are bounded in probability.}\end{array}
\eee
The noise $(\ep^n_i)_{i\geq0}$ can be realized as
$\ep^n_i=\chi_i$, where $(\chi_i)_{i\in\Z}$ is a stationary, centered
process, independent of the $\si$-field $\f_\infty=\vee_{t>0}\,\f_t$, and
with finite moments of all orders, and which is $v$-polynomially
$\rho$-mixing for some $v>1$.
\vsc

\nib Assumption (NO-2): \rm We have $T(n,i)=i\De_n$ (regular observation
scheme), and the noise $(\ep^n_i)_{i\geq0}$ can be realized as
\bee\label{S4}
\ep^n_i~=~\ga_{T(n,i)}\,\cdot \chi_i,
\eee
where $\ga$ is a nonnegative It\^o semimartingale on $\proba$, which
satisfies Assumption (H) (with of course different coefficients than
in (\ref{S1})), and $(\chi_i)_{i\in\Z}$ is as in (NO-1).
\vsc

\begin{rema}\label{RS1} \rm These assumptions could be weakened by asking
finite moments up to a suitable order only: for example, if one is
interested in estimating the covariance function of the process $\chi$,
we only need finite moments up to order $q$, bigger than but arbitrarily
close to $4$.

The $\rho$-mixing condition could also be replaced by $\al$-mixing or
$\phi$-mixing, or by any other condition implying ergodicity and a central
limit theorem for all functionals of the type $\sum_{i=1}^nf(\chi_i)$
when $\E(f(\chi_0))=0$ and $\E(|f(\chi_0)|^q)<\infty$ for all $q>0$.\qed
\end{rema}

\begin{rema}\label{RS2} \rm Under Assumption (NO-2) the noise is not really
independent
of $X$, a form of dependency being induced by the presence of the process
$\ga$. However (NO-2) and {\em a fortiori} (NO-1) imply that the noise and the
returns of $X$ are {\em not}\ correlated: this is of course a drawback of
the model used here.\qed
\end{rema}

\begin{rema}\label{RS3} \rm It should be noted that our model does {\em not}\
provide a definition of noise which is ``consistent'' with a change of
observation times, in the following sense: when $T(n,i)=i/n$ with $n$ even,
and when we subsample and take only the observations at times $2i/n$ (this
amounts to replacing $n$ by $n/2$), then in (\ref{S4}) we have to replace
the process $\chi$ by a new process $\chi'_i=\chi_{2i}$. This
new process shares the same mixing properties as $\chi$, but the covariance
is modified in a trivial way.\qed
\end{rema}

\section{Estimation of the moments of the noise}
\label{sec:M}
\setcounter{equation}{0}
\renewcommand{\theequation}{\thesection.\arabic{equation}}

We will be interested in estimating the various moments of the noise.
For this, we introduce some general notation: let $\ja$ be the set of all
finite sequences of relative integers
$\bj=(j_1,j_2,\cdots,j_q)$ (they are neither necessarily ordered, nor
necessarily distinct, and $q\geq1$), and we use the notation
\bee\label{M0}
\bj=(j_1,\cdots,j_q),~\bj'=(j'_1,\cdots,j'_{q'})~\rightsquigarrow~
\left\{\begin{array}{l}
q(\bj)=q,~~\mu(\bj)=
\max(j_1,\cdots,j_q)\\
\bj\oplus\bj'=(j_1,\cdots,j_q,j'_1,\cdots,j'_{q'})\\
\bj_{+m}=(m+j_1,\cdots,m+j_q)~~\text{if}~m\in\Z.
\end{array}\right.
\eee
We introduce the subset $\ja^+$ of $\ja$ consisting of all
$\bj=(j_1,j_2,\cdots,j_q)$ with $j_r\geq0$ for all $r=1,\cdots,q$,
and $\ja^{0+}$ is the set of all $\bj=(j_1,j_2,\cdots,j_q)\in\ja^+$ such
that $j_1=0$ and $q\geq2$.

Associated with each $\bj\in\ja$, we introduce the integer composite
moments of the noise $\chi$ as
\bee\label{M1}
R(\bj)=R(j_1,\cdots,j_q)=\E\Big(\prod_{r=1}^q\chi_{j_r}\Big)
~~~\text{if}~~\bj=(j_1,\cdots,j_q).
\eee
Note that $R(\bj)=0$ when $q(\bj)=1$, and $R(\bj)=R(\bj_{+m})$ for all
$m\in\Z$, so we restrict our attention to the estimation of $R(\bj)$ when
$\bj\in\ja^{0+}$. The covariance of $\chi$ is
$r(j)=R(0,j)$, the variance is $R(0,0)$.

\subsection{Consistency Results.}
For estimating $R(\bj)$ we first choose a sequence $k_n\geq2$ of integers
which satisfies, with $\De_n$ as in (\ref{S2}):
\bee\label{M2}
k_n~\to~\infty,\qquad
%%%
%k_n\,\De_n^\eta~\to~0,\quad \text{for some}~~\eta\in\big(0,\frac12\big).
k_n\,\De_n^\theta~\to~0,\quad \text{for some}~~\theta\in\big(0,\frac12\big).
%%%
\eee
Then we set
\bee\label{M3}
\BX^n_i=\frac1{k_n}\,\sum_{j=0}^{k_n-1} X_{T(n,i+j)},\quad
\BY^n_i=\frac1{k_n}\,\sum_{j=0}^{k_n-1} Y^n_{i+j},\quad
\Bep^n_i=\frac1{k_n}\,\sum_{j=0}^{k_n-1} \ep^n_{i+j},\quad
\Bchi^n_i=\frac1{k_n}\,\sum_{j=0}^{k_n-1} \chi_{i+j},
\eee
and for $\bj=(j_1,\cdots,j_q)$ and $\mu=\mu(\bj)$, consider the processes
\bee\label{M4}
U(\bj)^n_t=\sum_{i=0}^{N_n(t)+1-\mu-2qk_n}~
\prod_{r=1}^q(Y^n_{i+j_r}-\BY^n_{i+\mu+(2r-1)k_n}).
\eee
The index for $\BY^n_.$ above is chosen to ensure that the noise
components in $Y^n_{i+j_r}$ and in $\BY^n_{i+\mu+(2r-1)k_n}$ are separated
by at least $k_n$ indices, implying that they are ``independent enough''.
The sum above, as everywhere else below, is set to be $0$ when the upper
limit is smaller than  the lower limit, that is
$N_n(t)<
\mu+2qk_n-1$, but for any $t>0$ this is not the
case when $n$ is large enough. The upper limit of the sum above is
such that $U(\bj)_t$ uses only data within the time interval
$[0,t]$, and all these data.

The consistency results are as follows.

\begin{theo}\label{TN1} Assume (H) and (\ref{M2}). {Let $\bj\in\ja^+$
and $T>0$}.

(a) Under (NO-1) we have
\bee\label{N1}
\frac1{N_n(T)}\,U(\bj)^n_T~\toop~R(\bj).
\eee

(b) Under (NO-2) { we have
\bee\label{N2}
\De_n\,U(\bj)^n_T~\toop~R(\bj)\int_0^T\ga_s^{q(\bj)}\,ds.
\eee
}
\end{theo}

When both (NO-1) and (NO-2) hold, so
$\ga_t=1$, (a) is a special case of (b). Also, under (NO-2), there is a
fundamental non-identifiability, namely we
can divide $\ga$ by a number $a>0$, and multiply $\chi$ by
the same $a$: this explains the form of the limit in (\ref{N2}),
and in this case there is of course no way to estimate $R(\bj)$ any
better than up to a multiplicative constant.

In the next subsection we will state Central Limit Theorems associated
with these convergences. They involve some limiting variances-covariances,
based on the following quantities, where $\bj,\bj'\in\ja^+$:
\bee\label{M5}
\Si^{\bj,\bj'}=\sum_{m\in\Z}\big(R(\bj\oplus\bj'_{+m})
-R(\bj)\,R(\bj')\big),
\eee
and we will see in the proofs below that these are finite numbers,
and if $\ja_0$ is a finite subset of $\ja^+$ the matrix
$(\Si^{\bj,\bj'})_{\bj,\bj'\in\ja_0}$ is a covariance matrix.

In order to have ``feasible'' CLTs we need consistent estimators for
$\Si^{\bj,\bj'}$ in case of (NO-1), and for $\Si^{\bj,\bj'}\int_0^T
\ga_s^{q(\bj)+q(\bj')}\,ds$ in case of (NO-2). The previous theorem
gives us such consistent estimators in case of (NO-1), and consistent
estimators for $R(\bj\oplus\bj'_{+m})\int_0^T\ga_s^{q(\bj)+q(\bj')}\,ds$,
but not for $R(\bj)R(\bj')\int_0^T\ga_s^{q(\bj)+q(\bj')}\,ds$. For
estimating the latter quantity, we do as follows. If
$\bj=(j_1,\cdots,j_q)$ and $\bj'=(j'_1,\cdots,j'_{q'})$ are in $\ja^+$,
with $\mu=\mu(\bj)$ and $\mu''=\mu+\mu(\bj')$ and $q''=q+q'$, we set
\bee\label{N3}
\BU(\bj,\bj')^n_t=
\sum_{i=0}^{N_n(t)+1-\mu''-(2q''+1)k_n}
\prod_{r=1}^q(Y^n_{i+j_r}-\BY^n_{i+\mu+(2r-1)k_n})
\prod_{r=1}^{q'}(Y^n_{i+\mu+(2q+1)k_n+j'_r}-\BY^n_{i+\mu''+(2r+2q)k_n})
\eee
Then we have:

\begin{theo}\label{TN2} Assume (H), (NO-2) and (\ref{M2}), and let
$\bj,\bj'\in\ja^+$. If $k_n$ satisfies (\ref{M2}) {and $T>0$
we have
\bee\label{N5}
\De_n\,\BU(\bj,\bj')^n_T~\toop~R(\bj)R(\bj')
\int_0^T\ga_s^{q(\bj)+q(\bj')}\,ds.
\eee
}
\end{theo}

We have a similar result under (NO-1), but this is not needed below.
Coming back to the covariances $\Si^{\bj,\bj'}$, we have:

\begin{cor}\label{CN1} Assume (H) and (\ref{M2}). {Let $\bj,\bj'\in\ja^+$
and $T>0$}.

(a) Under (NO-1) and if $k'_n$ satisfies
\bee\label{N6}
k'_n~\to~\infty,\qquad k'_n\leq k_n,\qquad k'_n\,\sqrt{k_n\,\De_n}~\to~0,
\eee
then we have $\wSi^{\bj,\bj',n}_T\toop
\Si^{\bj,\bj'}$, where
\bee\label{N7}
\wSi^{\bj,\bj',n}_T=\frac1{N_n(T)}\Big(U(\bj\oplus\bj')^n_T
+2\sum_{m=1}^{k'_n}U((\bj\oplus\bj'_{+m})^n_T\Big)
-\frac{2k'_n+1}{N_n(T)^2}\,U(\bj)^n_T\,U(\bj')^n_T.
\eee

(b) Under (NO-2) and if $k_n$ satisfies (\ref{M2}) and $k'_n$ satisfies
\eqref{N6} we have $\wSi'^{\bj,\bj',n}_T\toop
\Si^{\bj,\bj'}\int_0^T\ga_s^{q(\bj)+q(\bj')}\,ds$, where
\bee\label{N8}
\wSi'^{\bj,\bj',n}_T=\De_n\Big(U(\bj\oplus\bj')^n_T
+\sum_{m=1}^{k'_n}(U((\bj\oplus\bj'_{+m})^n_T+U((\bj_{+m}\oplus\bj')^n_T
\Big)\Big)-(2k'_n+1)\De_n\,\BU(\bj,\bj')^n_T.
\eee
\end{cor}

\subsection{Central Limit Theorems.}

Suppose for instance that (NO-1) holds; by Theorem \ref{TN1} it looks like
$U(\bj)^n_T/N_n(T)$ properly estimates $R(\bj)$, but it turns out,
rather, that it is a good estimator for the following:
\bee\label{N4}
R(k_n;\bj)=\E\Big(\prod_{r=1}^{q(\bj)}\big(\chi_{j_r}
-\Bchi^n_{\mu(\bj)+(2r-1)k_n}\big)\Big),
\eee
and there is a CLT with this centering term under exactly the same
assumptions than in this theorem (and similarly when (NO-2) holds).
As far as consistency is concerned this is not a problem because
$R(k_n;\bj)\to R(\bj)$, as we will show below. For the CLT with the
desired centering $R(\bj)$, though, we need the convergence
$R(k_n;\bj)\to R(\bj)$ to be faster than
the rate of convergence, namely $\De_n^{1/2}$, in the CLT. This will
be the case if $k_n$ goes fast enough to $\infty$, and more precisely
if, instead of (\ref{M2}), we have the following, which is stronger
than (\ref{M2}), {and where $v$ is the mixing exponent}:
\bee\label{M12}
k_n\De_n^{1/2v}~\to~\infty,\quad
 k_n\De_n^{\theta}~\to~0\quad\text{for some}~~
 \theta\in\big(\frac1{2v},\frac12\big).
\eee

Below, we state two different theorems, under (NO-1) and (NO-2)
respectively, and $\ja_0$ is any finite subset of $\ja^+$. We also recall
that a sequence of variables $U_n$ on $(\Om,\f,\PP)$, taking their values in
some Polish space $E$, is said to converge $\f_\infty$-stably
in law to a limit $U$ defined on an extension $(\WOm,\Wf,\WP)$ of the
original space if, for any continuous bounded function $f$ on $E$ and
any bounded $\f_\infty$-measurable variable $\Psi$, we have
$\E(\Psi\,f(U_n))\to\WE(\Psi\,f(U))$.

\begin{theo}\label{TM1} Assume (H), (NO-1) and (\ref{M12}).
For any fixed $T>0$ the $\R^{\ja_0}$-valued random variables
$Z^n_T=(Z^{n,\bj}_T)_{\bj\in\ja_0}$ with components
\bee\label{M6}
Z^{n,\bj}_T~=~\sqrt{N_n(T)}\,\Big(\frac1{N_n(T)}~U(\bj)^n_T-R(\bj)\Big)
\eee
converge $\f_\infty$-stably in law to a centered Gaussian $\R^{\ja_0}$-valued
variable defined on an extension $(\WOm,\Wf,\WP)$ of the space, independent
of $\f_\infty$,
and whose covariance matrix is $\Si^{\bj,\bj'}$, as defined by (\ref{M5}).
\end{theo}

In this result, similar with (a) of Theorem \ref{TN1} and in contrast with
the next result to come, we do {\em not} have the functional convergence (as
processes), and the limit is not even depending on $T$.

Note also that we could as well consider the whole (countable) family $\ja^+$
instead of a finite subset $\ja_0$, if we consider the product topology
on $\R^{\ia}$. The same comment applies to the forthcoming
result as well, but we do not need this kind of generality in this paper.

\begin{theo}\label{TM2} Assume (H) and (NO-2) and (\ref{M2}). {For any fixed
$T>0$ the $\R^{\ja_0}$-valued random variables
$Z'^n_T=(Z'^{n,\bj}_T)_{\bj\in\ja_0}$ with
components
\bee\label{M7}
Z'^{n,\bj}_T~=~\rdnn\,\Big(\De_n\,U(\bj)^n_T
-R(\bj)\int_0^t\ga_s^{q(\bj)}\,ds\Big)
\eee
converge $\f_\infty$-stably in law to a variable
$Z'_T=(Z'(\bj)_T)_{\bj\in\ja_0}$, defined on an extension $(\WOm,\Wf,\WP)$ of
the space, which,
conditionally on $\f_\infty$, is centered Gaussian with
(conditional) covariance
\bee\label{N9}
\E\big(Z'^{\bj}_T\,Z'^{\bj'}_T\mid\f_\infty)~=~\Si^{\bj,\bj'}
\int_0^T\ga_s^{q(\bj)+q(\bj')}\,ds.
\eee
}
\end{theo}

These results, joint with Corollary \ref{CN1}, also give us feasible
CLTs, in the following sense: suppose for simplicity that $\ja_0=\{\bj\}$
is a singleton. Then with notation (\ref{N8}), we have
\bee\label{N10}
\frac1{\sqrt{\De_n\,\wSi_T^{\bj,\bj,n}}}\,\Big(\De_n\,U(\bj)^n_T
-R(\bj)\int_0^T\ga_s^{q(\bj)}\,ds\Big)~\tol~\n(0,1)
\eee
under the assumptions of Theorem \ref{TM2} (we even have the
$\f_\infty$-stable convergence in law above). This is due, by standard
properties of the stable convergence in law, to the fact the limit in
probability of $\wSi_T'^{\bj,\bj,n}$ is an $\f_\infty$-measurable variable.
In the same way, under the assumptions of Theorem \ref{TM1},
\bee\label{N11}
\sqrt{\frac{N_n(T)}{\wSi_T^{\bj,\bj,n}}}\,\Big(\frac1{N_n(T)}~U(\bj)^n_T
-R(\bj)\Big)~\tol~\n(0,1)
\eee

\begin{rema}\label{RN1} \rm The choice of $k_n$ in (\ref{M12}) requires
the knowledge of $v$, or at least of the fact that $v$ is bigger
than some known value $v'>1$: in this case one may take $k_n\asymp
1/\De_n^{1/2v'}$. This is unfortunate, since in general one
does not {\em a priori}\ know the law of $\chi$, and in particular whether
it is stationary, or mixing, not to mention the number $v$ for which
it is $v$-polynomially $\rho$-mixing.
Nevertheless, nothing can be done without assumptions, and assuming
that the unknown $v$ is bigger than some fixed $v'>1$ seems reasonably weak.
In practice, one can choose $k_n$ in an
ad-hoc manner: first pick a preliminary $k_n$ and check how fast the estimated
correlations decay. If the estimated correlations decay fast, then
one can possibly switch to a smaller $k_n$, otherwise one may increase $k_n$.
One can also get some guidance from simulation studies, by coining a
time series whose auto-correlations has similar behavior to what is observed
in the real data, and then choosing different values of~$k_n$ in the
simulation study to see which values of  $k_n$ work better.
\end{rema}

\subsection{Estimation of the Covariance and the Correlation under (NO-1).}

In this subsection we assume (NO-1) and of course (H). A natural estimator
for the covariance $r(j)$ for any given $j\geq0$ is as follows (the time
horizon $T$ is fixed): we choose two sequences $k_n$ and $k'_n$ satisfying
(\ref{M12}) and (\ref{N6}), respectively, and set
\bee\label{MA1}
\wwr(j)_n~=~\frac1{N_n(T)}~U(0,j)^n_T.
\eee
These estimators are consistent for estimating $r(j)$, and enjoy a Central
Limit Theorem with rate $1/\sqrt{N_n(T)}$ and an asymptotic variance which
is consistently estimated by
$$\wSi(j)_T^n=\frac1{N_n(T)}\,\Big(U(0,j,0,j)^n_T
+2\sum_{m=1}^{k'_n}U(0,j,m,j+m)^n_T\Big)-(2k'_n+1)(\wwr(j)_n)^2.$$
Then we can rewrite (\ref{N11}) in this special case as
\bee\label{MA5}
\sqrt{\frac{N_n(T)}{\wSi(j)_T^n}}~(\wwr(j)_n-r(j)\big)~\tol~\n(0,1),
\eee
and, recalling that $N_n(T)$ is obviously
known to the statistician, it is straightforward to construct
confidence intervals for any $r(j)$.

\begin{rema}\label{RMA1} \rm (\ref{MA5}) is not the end of the story
about covariance estimation. Since Theorem \ref{TM1} is a multivariate
result, it is no problem to find confidence
bounds for any finite family $(r(0),r(1),\cdots,r(j))$, although
the formulation becomes messier when $j$ increases. \qed
\end{rema}

\begin{rema}\label{RMA2} \rm The choice of both sequences $k_n$ and $k'_n$
is connected with the numbers $\De_n$ in (NO-1), which are a kind of
mesh sizes, and also with the number $v$. The -- annoying -- connection
with $v$ has been discussed in Remark \ref{RN1}.
The connection with $\De_n$ is even more annoying, in a sense: under (NO-1),
these numbers $\De_n$ are unknown (or,
unobservable), although they are supposed to exist.

However, although we do not develop this topic here in a formal way,
it can be shown that good proxies for $\De_n$ are the observable
numbers $T/N_n(T)$. Indeed, we can replace (\ref{M12}) and (\ref{N6})
by
\bee\label{MA601}
%%%
%k_n\asymp N_n(T)^\te,\quad k'_n\asymp N_n(T)^{\te'},\quad\text{\rm where}~~
%\frac1{2v}<\te<\frac12,~~\te'<1-\te.
k_n\asymp N_n(T)^\eta,\quad k'_n\asymp N_n(T)^{\eta'},\quad\text{\rm where}~~
\frac1{2v}<\eta<\frac12,~~\eta'<\frac{1-\eta}{2} \wedge \eta.
%1-\eta.
%%%
\eee
Then, although $k_n$ and $k'_n$ are now random, all the previous
results still hold, if (NO-1) holds for a possibly unknown sequence
$\De_n\to0$: this is due to the fact that, since we are here analyzing the
noise with the structure $\ep^n_i=\chi_i$, the calendar time is
relatively of little importance in comparison with the index $i$
enumerating the observations themselves.\qed
\end{rema}

\begin{rema}\label{RMA3} \rm If one does as suggested in the previous remark,
one is still left with the important problem of choosing the tuning
parameters $\eta$ and $\eta'$, and also the proper proportionality constants.
This is exactly
as in all statistical problems for which one uses local windows.\qed
\end{rema}

Now, we turn to the estimation of the correlation between $\chi_i$ and
$\chi_{i+j}$, for a fixed $j\geq1$, that is
\bee\label{MA6}
\Co(j)~=~\frac{r(j)}{r(0)}.
\eee
Using (\ref{MA1}), natural (and consistent) estimators for this are
\bee\label{MA7}
\wCo(j)_n~=~\frac{\wwr(j)_n}{\wwr(0)_n}.
\eee

The associated CLT is a straightforward consequence of Theorem \ref{TM1},
%%%%-(a),
used with $\bj=(0,0)$ and
$\bj'=(0,j)$. Namely, under (H) and (NO-1) and (\ref{M12}) for $k_n$,
$\sqrt{N_n(T)}\,\big(\wCo(j)_n-\Co(j)\big)$ converges $\f_\infty$-stably
in law to a variable which is $\f_\infty$-conditionally centered Gaussian
with (conditional) variance
$$\s(j)~=~\frac{r(0)^2\,\Si^{(0,j),(0,j)}+r(j)^2\,\Si^{(0,0),(0,0)}
-2r(0)r(j)\,\Si^{(0,0),(0,j)}}{r(0)^4}.$$
With the notation (\ref{N7}), consistent estimators for $\s(j)$ are
$$\ws(j)^n_T=\frac1{\wwr(0)_n^4}\,\big(\wwr(0)_n^2\,\wSi_T^{(0,j),(0,j),n}
+\wwr(j)_n^2\,\wSi_T^{(0,0),(0,0),n}
-2\wwr(j)_n\,\wwr(0)_n\,\wSi_T^{(0,0),(0,j),n}\big).$$

At this stage, the following result is obvious:

\begin{theo}\label{TA1} Assume (H) and (NO-1), and let $k_n$
satisfy (\ref{M12}) and $k'_n$ satisfy (\ref{N6}). Then
\bee\label{cor1}
\sqrt{\frac{N_n(T)}{\ws(j)_T^n}}~(\wCo(j)_n-\Co(j)\big)~\tol~\n(0,1).
\eee
\end{theo}

\subsection{Estimation of the Covariance and the Correlation under (NO-2).}

From now on, we suppose that we are under (NO-2), with regularly spaced
observations and the additional process $\ga$. As mentioned before, we
cannot estimate the covariance $r(j)$, but we can estimate the
``integrated covariance'', which is
\bee\label{MA2}
\ra(j)_T~=~r(j)\,\int_0^T\ga^2_s\,ds.
\eee
Or, perhaps, one would like to estimate the ``averaged'' observed
covariance $\ra(j)_T/T$ (this is of course the same problem), or the
``spot'' covariance $r(j)\ga_t^2$ at some time $t$ within $[0,T]$.

Despite its interest, we will not speak about spot covariance here,
since this is somewhat similar (because $\ga$ is supposed to be a
semimartingale) to the estimation of the spot volatility. The
estimation of $\ra(j)_T$ here is exactly the same problem
as the estimation of $r(j)$ under (NO-1): we choose two sequences $k_n$ and
$k'_n$ satisfying (\ref{M12}) and (\ref{N6}) respectively, and
a natural sequence of consistent estimators is given by
\bee\label{MA12}
\wra(j)^n_T~=~\De_n\,U(0,j)^n_T.
\eee
The rate of convergence in the CLT is now $\rdn$ and, recalling the
notation (\ref{N3}), the asymptotic variance is consistently estimated by
$$\wSi'(j)_T^n=\De_n\,\Big(U(0,j,0,j)^n_T+2\sum_{m=1}^{k'_n}
U(0,j,m,j+m)^n_T\Big)-(2k'_n+1)\De_n\,\BU((0,j),(0,j))^n_T.$$
Then we can rewrite (\ref{N10}) as
\bee\label{MA15}
\frac1{\sqrt{\De_n\,\wSi'(j)_T^n}}~(\wra(j)_T^n-\ra(j)_T\big)
~\tol~\n(0,1).
\eee

More interesting perhaps, in this case, is the estimation of the correlation
$\Co(j)$, which is given by (\ref{MA6}) but also satisfies
$\Co(j)=\frac{\ra(j)_t}{\ra(0)_t}$
for all $t>0$ (recall that $\ga>0$). Thus, consistent
estimators for this are {
\bee\label{MA17}
\wCo'(j)_T^n~=~\frac{\wra(j)_T^n}{\wra(0)_T^n}.
\eee
Again, $\rdnn\,\big(\wCo'(j)_T^n-\Co(j)\big)$ converges $\f_\infty$-stably
in law} to a variable which conditionally on
$\f_\infty$ is centered normal with variance $\s'(j)_T$ given by
$$\s'(j)_T~=~\frac{\ra(0)_T^2\,\Si^{(0,j),(0,j)}+\ra(j)_T^2\,\Si^{(0,0),(0,0)}
-2\ra(0)_T\ra(j)_T\,\Si^{(0,0),(0,j)}}{\ra(0)_T^4}\,\int_0^T\ga_s^4\,ds,$$
which is also equal to $\s(j)\int_0^T\ga_s^4\,ds/
\big(\int_0^T\ga_s^2\,ds\big)^2$.
Then, consistent estimators for $\s'(j)_T$ are
$$\ws'(j)^n_T=\frac1{(\wra(0)_T^n)^4}\,\big(
(\wra(0)_T^n)^2\,\wSi_T'^{(0,j),(0,j),n}
+(\wra(j)_T^n)^2\,\wSi_T'^{(0,0),(0,0),n}
-2\wra(j)_T^n\,\wra(0)_t^n\,\wSi_T'^{(0,0),(0,j),n}\big),$$
and we have the following:

\begin{theo}\label{TA2} Assume (H) and (NO-2), and let $k_n$
satisfy (\ref{M12}) and $k'_n$ satisfy \eqref{N6} {and $T>0$. Then
\bee\label{cor2}\frac1{\sqrt{\De_n\,\ws'(j)_T^n}}~(\wCo'(j)_T^n-\Co(j)\big)
~\tol~\n(0,1).\eee
}
\end{theo}

\section{Simulations}\label{sec:Sim}

%\subsection{Part I: Estimating the moments of noise}
Throughout the following two sections we take $T=1$, in other words, we
concentrate on intraday data.

\subsection{Under Assumption (NO-1)}
We consider the following design:
$X$ is an Ornstein-Uhlenbeck process with jumps
\bee\label{sim_X}
  dX_t = -\rho (X_t - \mu) + \sigma \ dW_t +  dJ_t,
\eee
where $W_t$ is a standard Brownian motion, and $J_t$ is a compound Poisson
process independent of $W$ as follows
\[
   J_t = \sum_{i=0}^{N_t} D_i,
\]
where $N_t$ is a Poisson process with rate $\lambda$, and $D_i$'s are
\mbox{i.i.d.} symmetric mixed normals:
$D_i = B_i \cdot Z_i' $, where $B_i$ takes values $1$ and $-1$ with equal
probability $0.5$, and $Z_i'\sim N(\mu',\sigma'^2)$.
The observation times $T(n,i)$ are specified as a Poisson process with rate
$n$, and $\chi$ is an AR$(1)$ process:
\[
  \chi_{i+1} = \phi\chi_i + Z_{i+1}^{''},\quad\mbox{where } Z_{i}^{''}
  \sim_{i.i.d.} N(0,\sigma_0^2).
\]
The observations are $Y^n_i = X_{T(n,i)} + \ep^n_i= X_{T(n,i)} + \chi_i.$

The specification of parameters is as follows:
\bee\label{par_Xchi}
\aligned
  \rho&=0.5, \q\mu=0.002, \q\sigma = 0.01, \q \lambda = 3,\q \mu'=\sigma/10,
  \q \sigma'=\sigma/30,\\
  \phi&=0.8,\quad \sigma_0=0.0003, \q \mbox{and}\q  n = 93,600.
  \endaligned
\eee

\begin{figure}[H]
\begin{center}
\includegraphics[height=6cm]{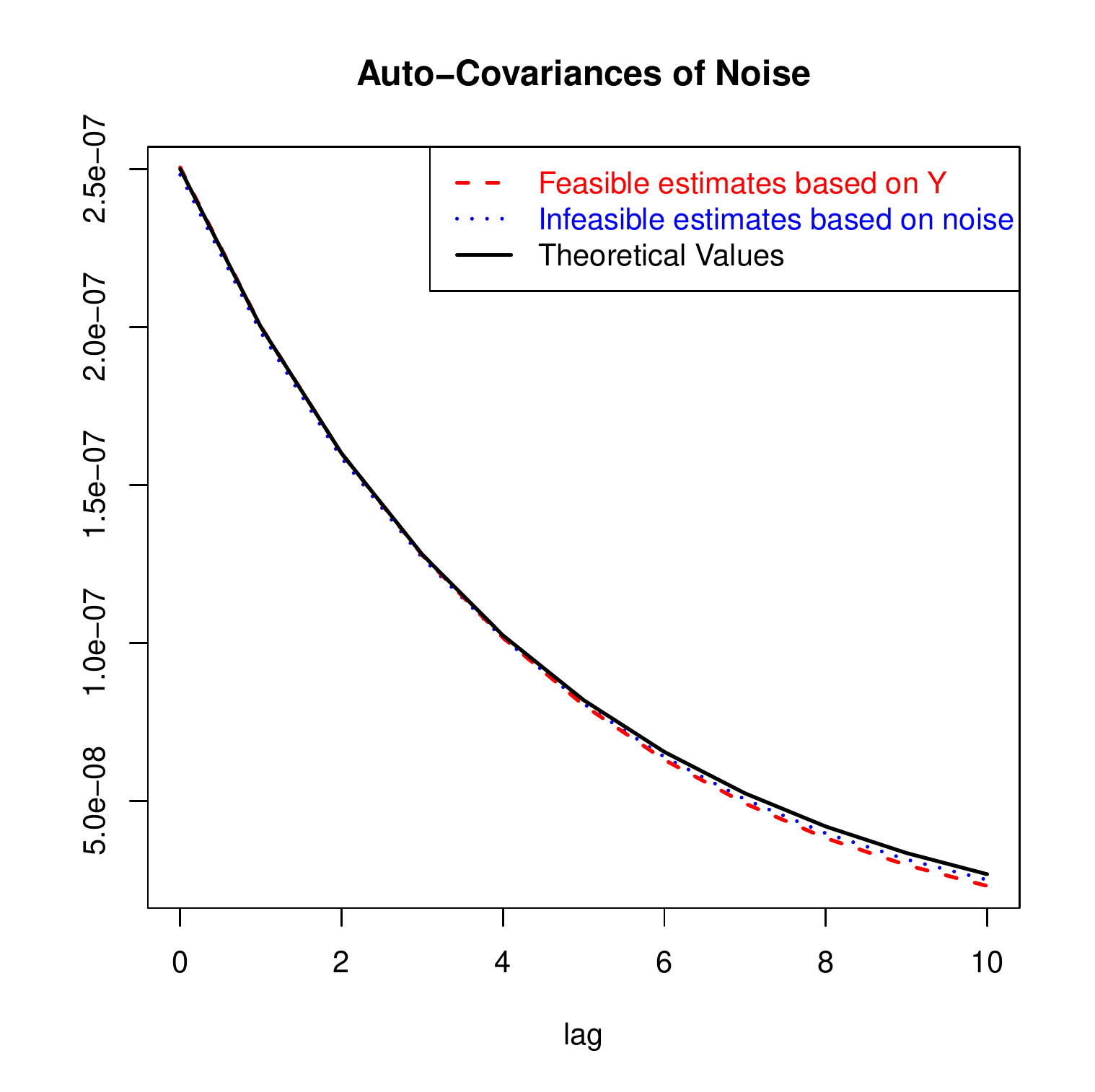}
\includegraphics[height=6cm]{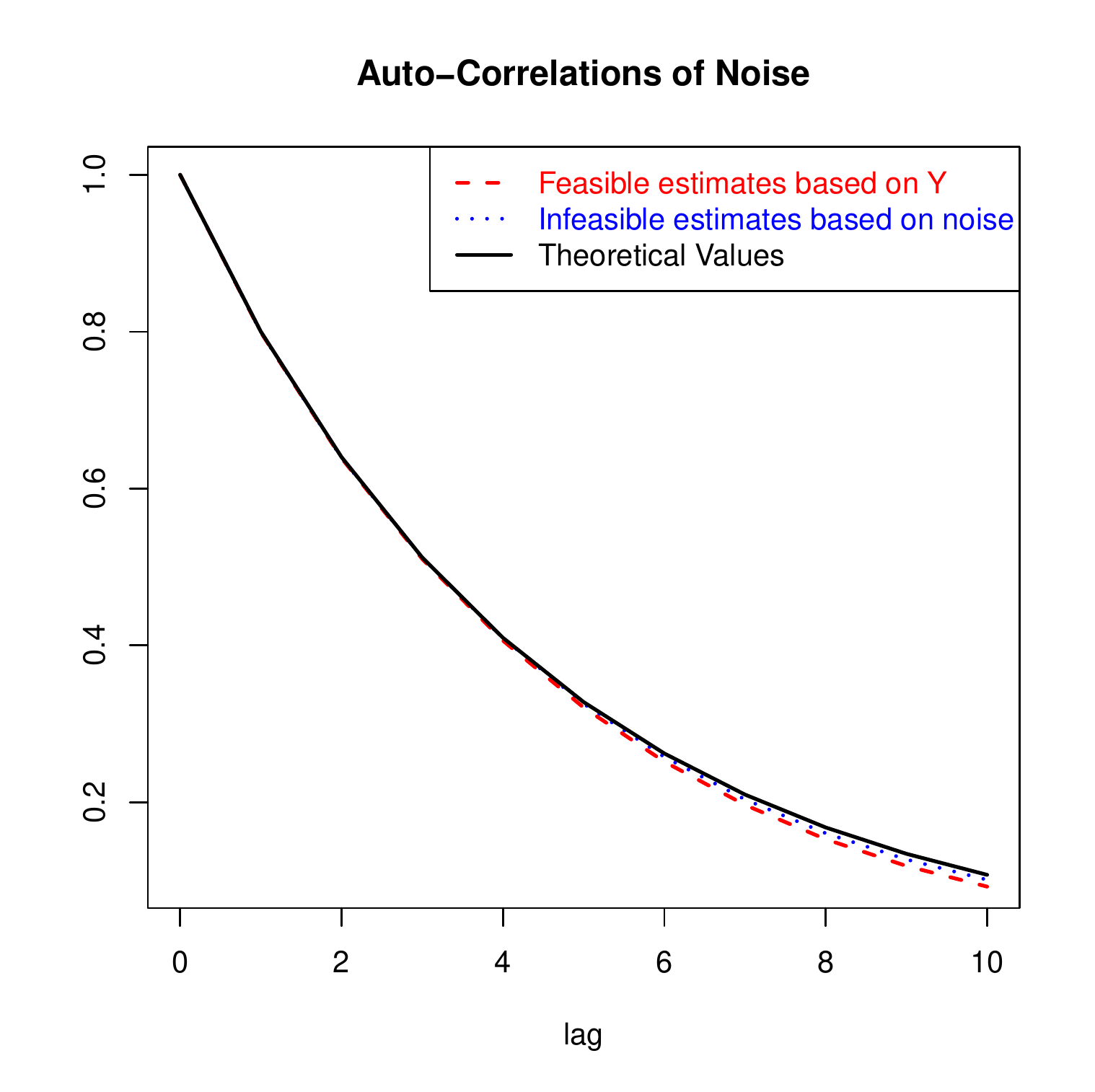}
\end{center}
  \caption{Comparison of estimates of auto-covariances and auto-correlations.}
\label{Fig:est_No1}
\end{figure}

In the estimation, we choose $k_n=8$ and $k_n'=4$.  Figure \ref{Fig:est_No1}
compares the estimates of auto-covariances and auto-correlations based on
Theorems \ref{TN1} and \ref{TM1} with the infeasible estimates based on the
noise process and the theoretical values. The estimates are  based on one
simulated path. More specifically, in Figure \ref{Fig:est_No1}, the dashed
red curves report $\wwr(j)_n$ as in \eqref{MA1} on the left and $\wCo(j)_n$
as in \eqref{MA7} on the right; the dotted blue curves report the
auto-covariance on the left  and the  auto-correlation on the right based on
the simulated $\chi$; the solid black curves report the theoretical values,
i.e.,
\bee\label{tv_sim}
r(j)= \frac{\phi^j \sigma_0^2}{1-\phi^2}, \q\mbox{and}\q \Co(j) =
\phi^j,\q\mbox{for } j\in \Z.
\eee
 Figure \ref{Fig:est_No1} demonstrates that under Assumption (NO-1), our
 estimates are comparable to the infeasible estimates based on the noise
 process, which is the best one can hope for; both are almost
 indistinguishable from the theoretical values.

\begin{figure}[H]
\begin{center}
 \includegraphics[height=8cm]{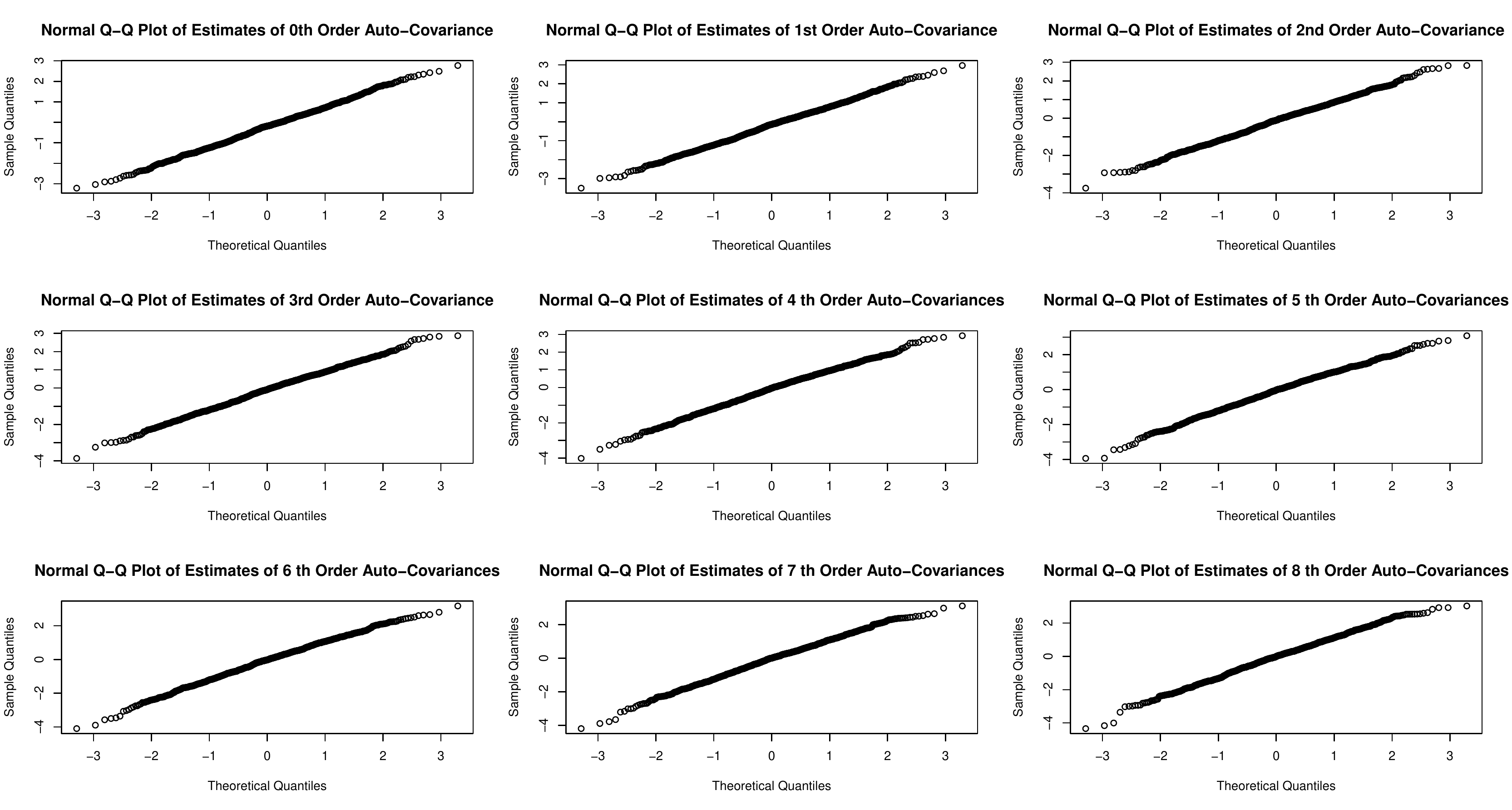}
   \end{center}
  \caption{QQ-plots of estimates of auto-covariances of orders from 0 to 8,
  based on 1,000 replications}
  \label{Fig:QQ_Cov_No1}
\end{figure}

\begin{figure}[H]
\begin{center}
  \includegraphics[height=8cm]{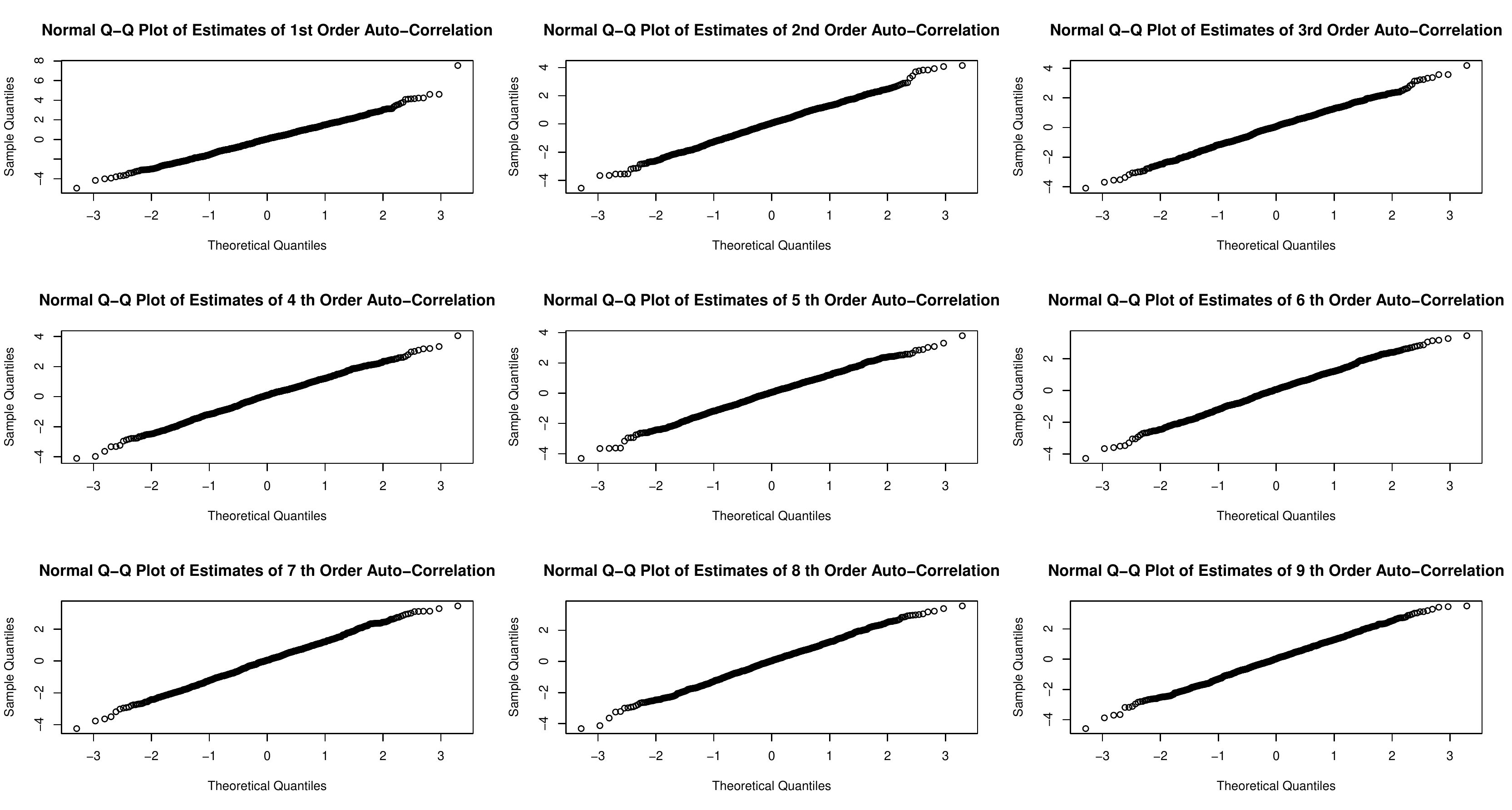}
\end{center}
 \caption{QQ-plots of estimates of auto-correlations of orders from 1 to 9, based on 1,000 replications}
 \label{Fig:QQ_Corr_No1}
\end{figure}

Next we demonstrate the central limit theorems  Theorem \ref{TM1} and
 Theorem \ref{TA1}.   We plot the normal quantile-quantile plots of
 $$\sqrt{\frac{N_n(T)}{\wSi_T^{\bj,\bj,n}}}\,\Big(\frac1{N_n(T)}~U(\bj)^n_T
-R(\bj)\Big)$$ as in (\ref{N11}) for $j=0,\cdots,8$ in Figure
\ref{Fig:QQ_Cov_No1}, and
 $$\sqrt{\frac{N_n(T)}{\ws(j)_T^n}}~(\wCo(j)_n-\Co(j)\big)$$
as in (\ref{cor1}) for $j=1,\cdots,9$ in Figure \ref{Fig:QQ_Corr_No1}, based
on 1,000 replications. All the plots support that the normality established
in the theorems can be relied on in practice with sample observed at a
reasonably high frequency within the time period being considered.

\subsection{Under Assumption (NO-2)}

The $X$ process is taken to be the same as in (\ref{sim_X}) above, namely,
an Ornstein-Uhlenbeck process with jumps. Under Assumption (NO-2), we have
an additional process $\gamma$, which we  assume to be an Ornstein-Uhlenbeck
process
\[
  d\gamma_t = -\rho_{\gamma} (\gamma_t - \mu_{\gamma}) + \sigma_{\gamma}
  \ dW_t,
\]
where $W_t$ is the \emph{same} Brownian motion that is used in (\ref{sim_X}).
$\chi$ is again an AR$(1)$ process. The observations are $Y^n_i = X_{T(n,i)}
+ \ep^n_i= X_{T(n,i)} + \gamma_{T(n,i)}\chi_i.$ Note that in this case the
noise $\ep^n_i$ is \emph{dependent} on the $X$ process.

The parameters for $X$, i.e., $\rho,\mu,\sigma,\lambda,\mu', \sigma', \phi$
and $\chi$ are the same as in \eqref{par_Xchi}. The parameters for $\gamma$
are $\rho_{\gamma}=0.5, \mu_{\gamma} =1 $ and $\sigma_{\gamma} = 0.01$. We
further take $\De_n=1/n:=1/93,600$.

\begin{figure}[H]
\begin{center}
\includegraphics[height=5cm]{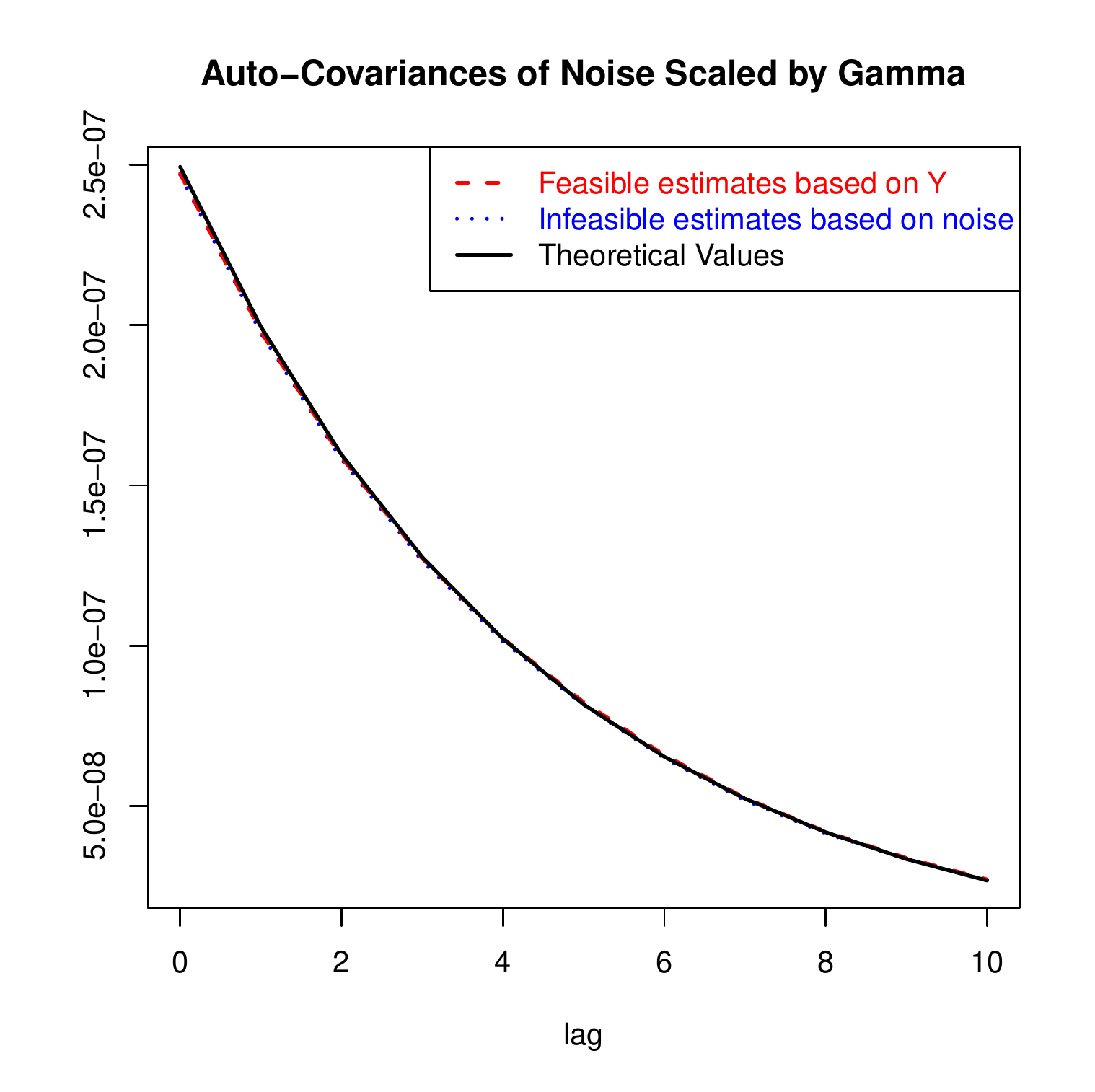}
\includegraphics[height=5cm]{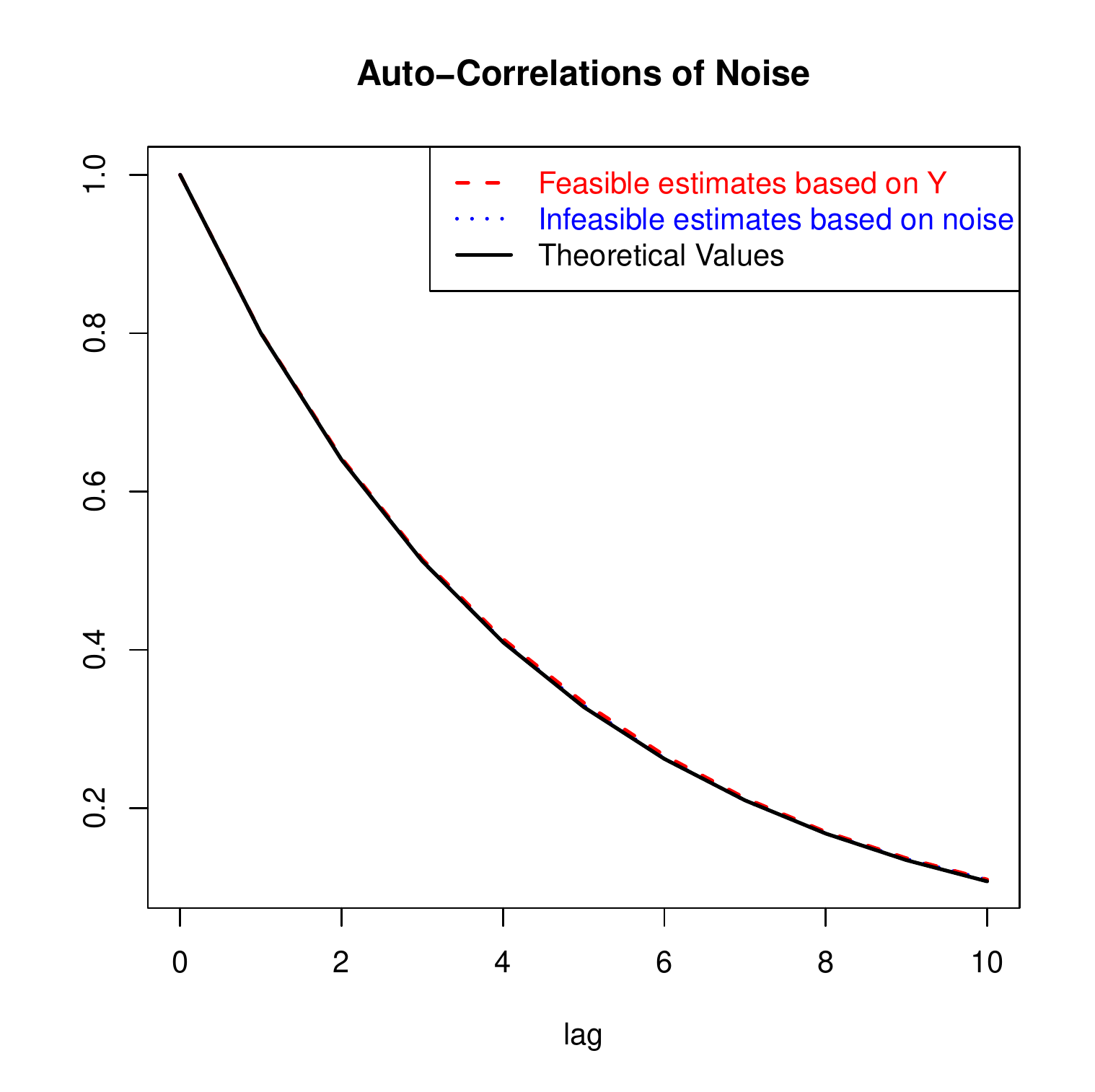}
\end{center}
  \caption{Comparison of estimates of auto-covariances and auto-correlations.}
\label{Fig:est_No2}
\end{figure}

In the estimation, we choose $k_n=8$ and $k_n'=4$. Figure \ref{Fig:est_No2}
compares the feasible estimates with the infeasible estimates and the
theoretical values. The estimates are again based on one simulated path.
More specifically, on the left panel, we use red dashed curve to report the
feasible estimates of the (scaled)  auto-covariances based on Theorem
\ref{TN1} ($\wra(j)^n$ as in \eqref{MA12}); blue dotted curve to report
the   infeasible estimates based on the noise process
$\ep^n_i=\gamma_{i\De_n} \chi_i$ (i.e., the  auto-covariance
based on the simulated $\chi$ multiplied by
$\widehat{\int_0^1\gamma_s^2\ ds} = \De_n\sum_{i=0}^n \gamma_{i\De_n}^2$);
black solid curve to report the theoretical values, i.e., $\ra(j)$ as in
\eqref{MA2}. On the right panel, we compare the feasible estimates of
auto-correlations based on Theorem \ref{TM2} ($\wCo(j)^n$ as in
\eqref{MA17}; see red dashed curve) with the infeasible estimates
based on the noise process (auto-correlations based on the simulated
$\chi$; see blue dotted curve), and the theoretical values ($\Co(j)$ as in
\eqref{tv_sim}; see black solid curve).
From Figure \ref{Fig:est_No2} we see again that our estimates are comparable
to the infeasible estimates based on the noise process, and  both are very
close to the theoretical values.

\begin{figure}[H]
\begin{center}
\includegraphics[height=7cm]{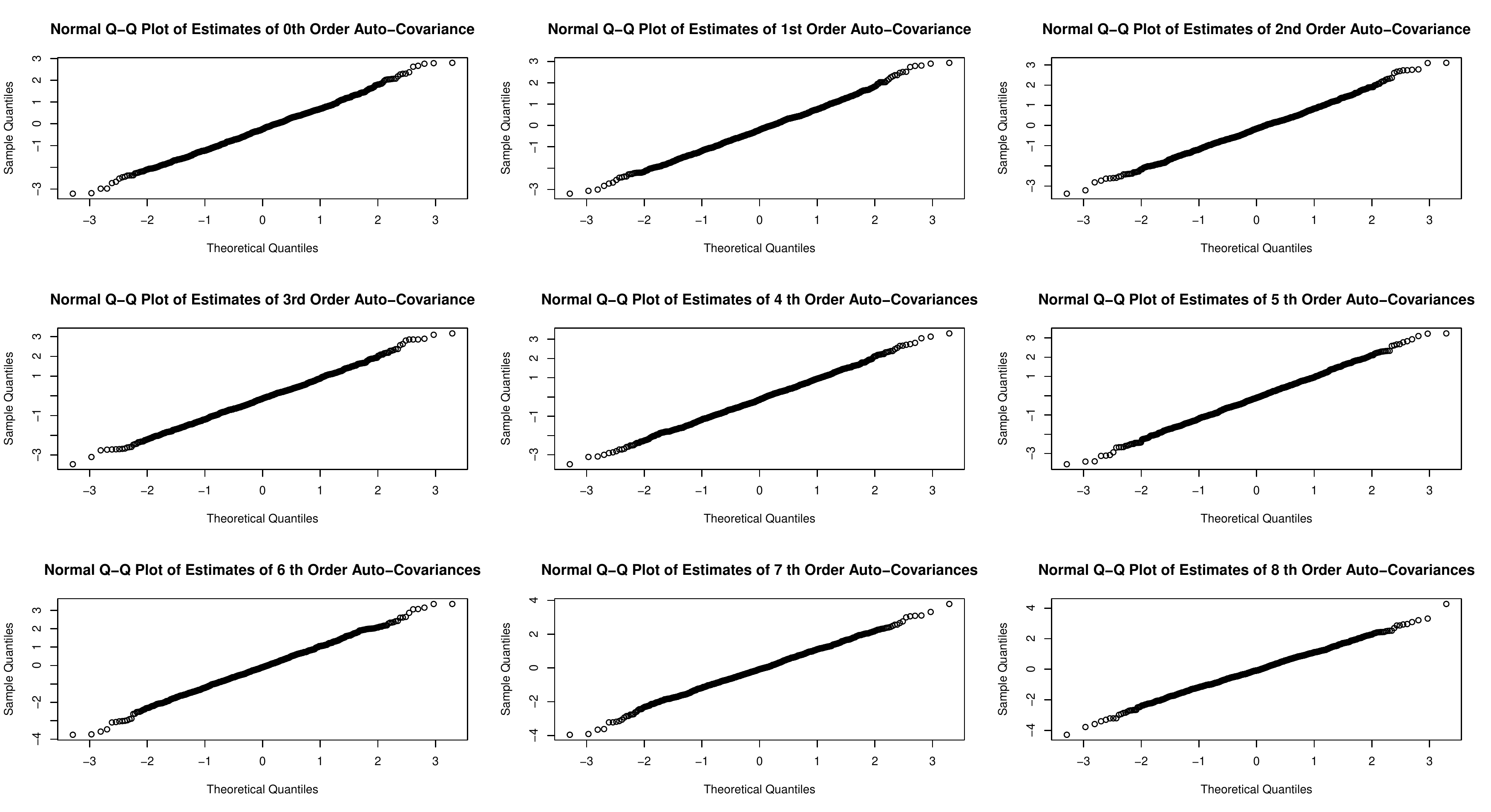}
\end{center}
     \caption{QQ-plots of estimates of auto-covariances of
     orders from 0 to 8, based on 1,000 replications}
\label{Fig:QQ_Cov_No2}
\end{figure}

\begin{figure}[H]
\begin{center}
\includegraphics[height=7cm]{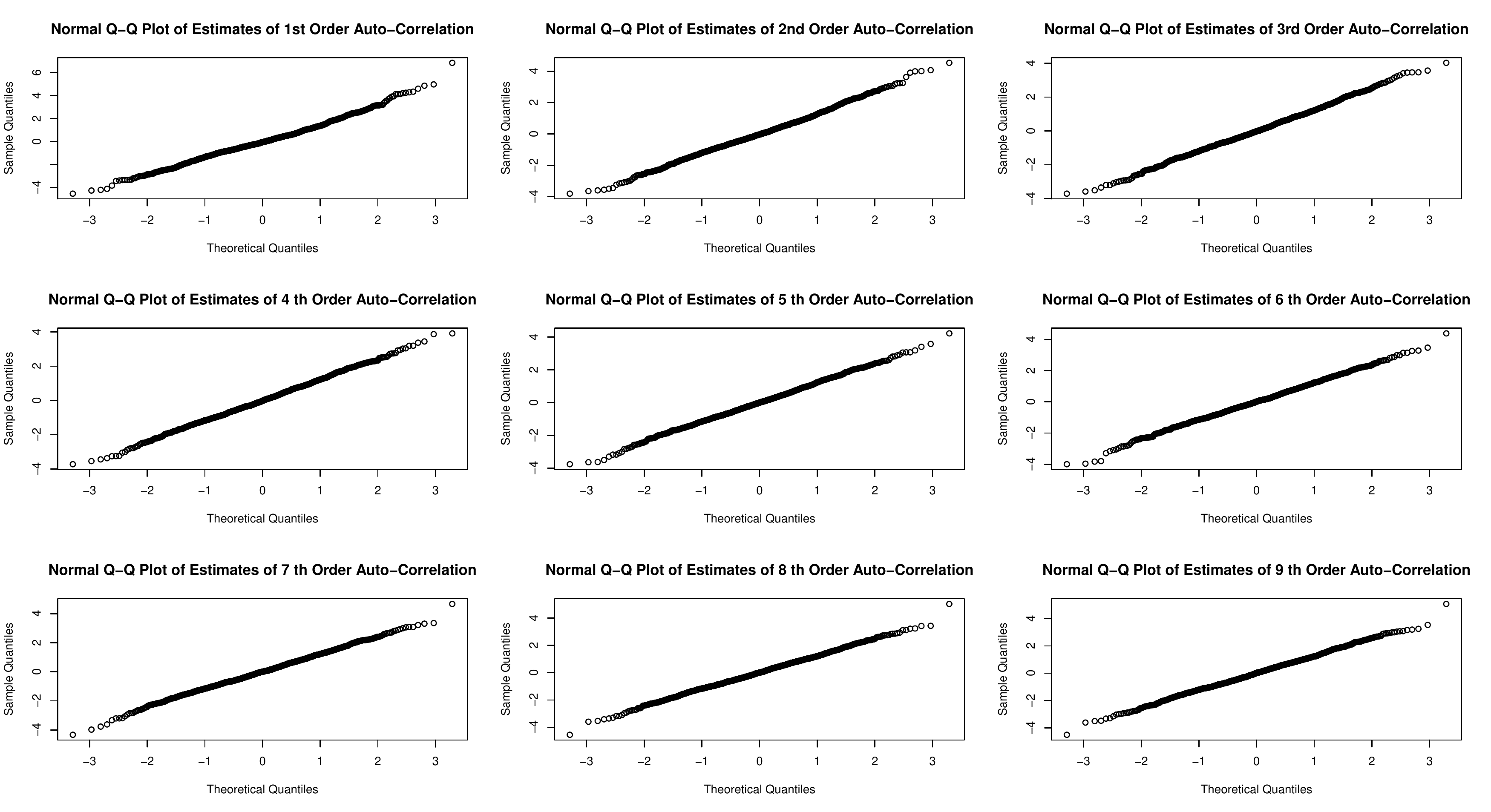}
\end{center}
\caption{QQ-plots of estimates of auto-correlations of orders from 1 to 9,
based on 1,000 replications}
\label{Fig:QQ_Corr_No2}
\end{figure}

Next we demonstrate the central limit theorems Theorem \ref{TM2} and Theorem
\ref{TA2}. The normal quantile to quantile plots of
\[
\frac1{\sqrt{\De_n\,\wSi_T^{\bj,\bj,n}}}\,\Big(\De_n\,U(\bj)^n_T -R(\bj)
\int_0^T\ga_s^{q(\bj)}\,ds\Big)
\]
 as in (\ref{N10}) for $j=0,\cdots,8$ are plotted in Figure
 \ref{Fig:QQ_Cov_No2} and the normal quantile to quantile plots of
$$\frac1{\sqrt{\De_n\,\ws'(j)_T^n}}~(\wCo'(j)_T^n-\Co(j)\big)$$ as in
(\ref{cor2})
for $j=1,\cdots, 9$ are plotted in Figure \ref{Fig:QQ_Corr_No2}, based on
1,000 replications.
Again, the practical applicability of the established normality is strongly
supported.

\section{Empirical Studies}\label{sec:emp}

In this section we examine the dependence of the microstructure noise for
several financial stocks, in particular, we estimate the auto-covariances
and auto-correlations and test whether they are equal to zero, based on
Theorems \ref{TM1} and \ref{TA1}.

\subsection{Citi Jan 2011 Data}
\begin{figure}[H]
\begin{center}
  \includegraphics[height=6.2cm]{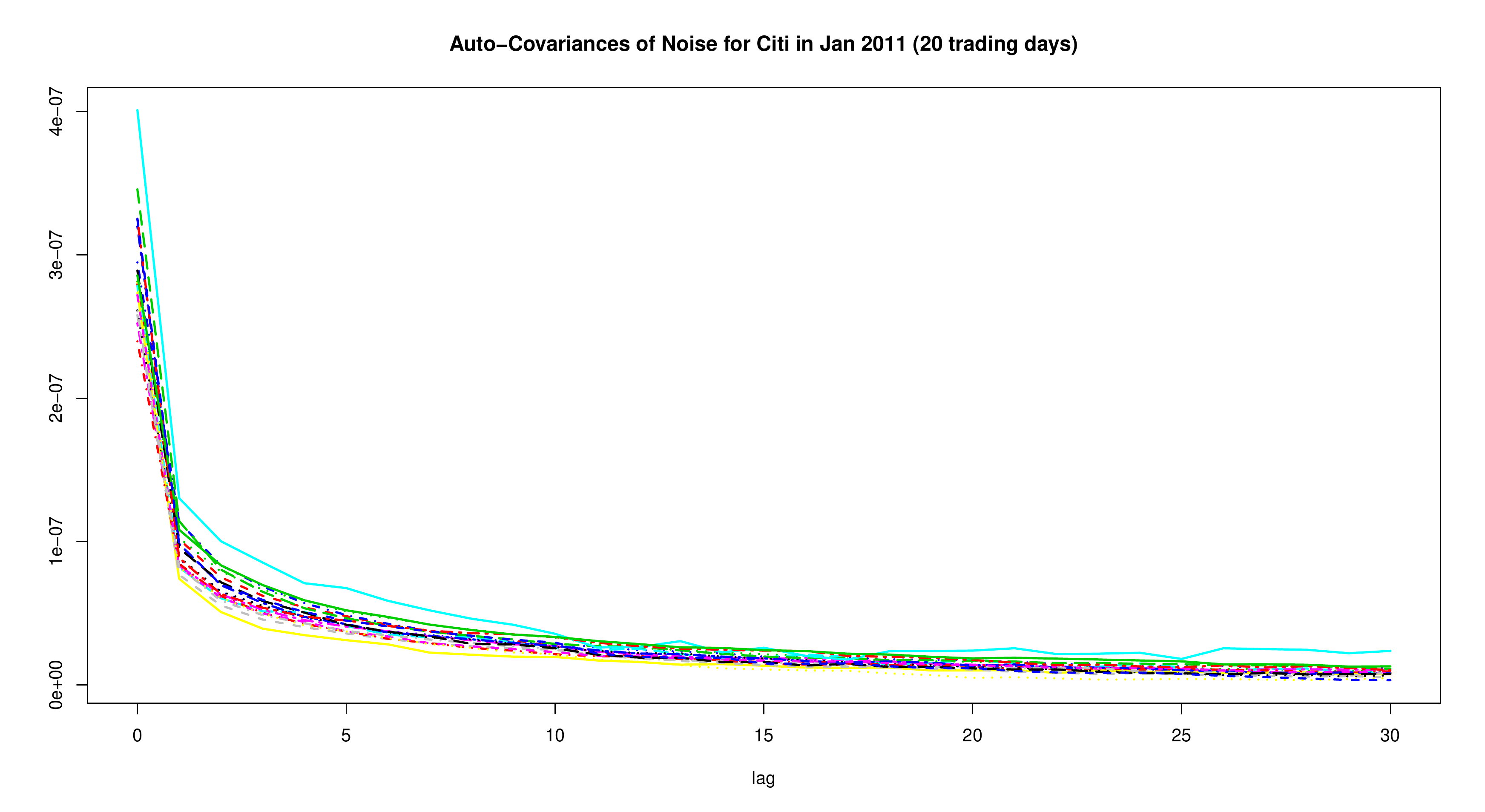}\\
  \includegraphics[height=6.2cm]{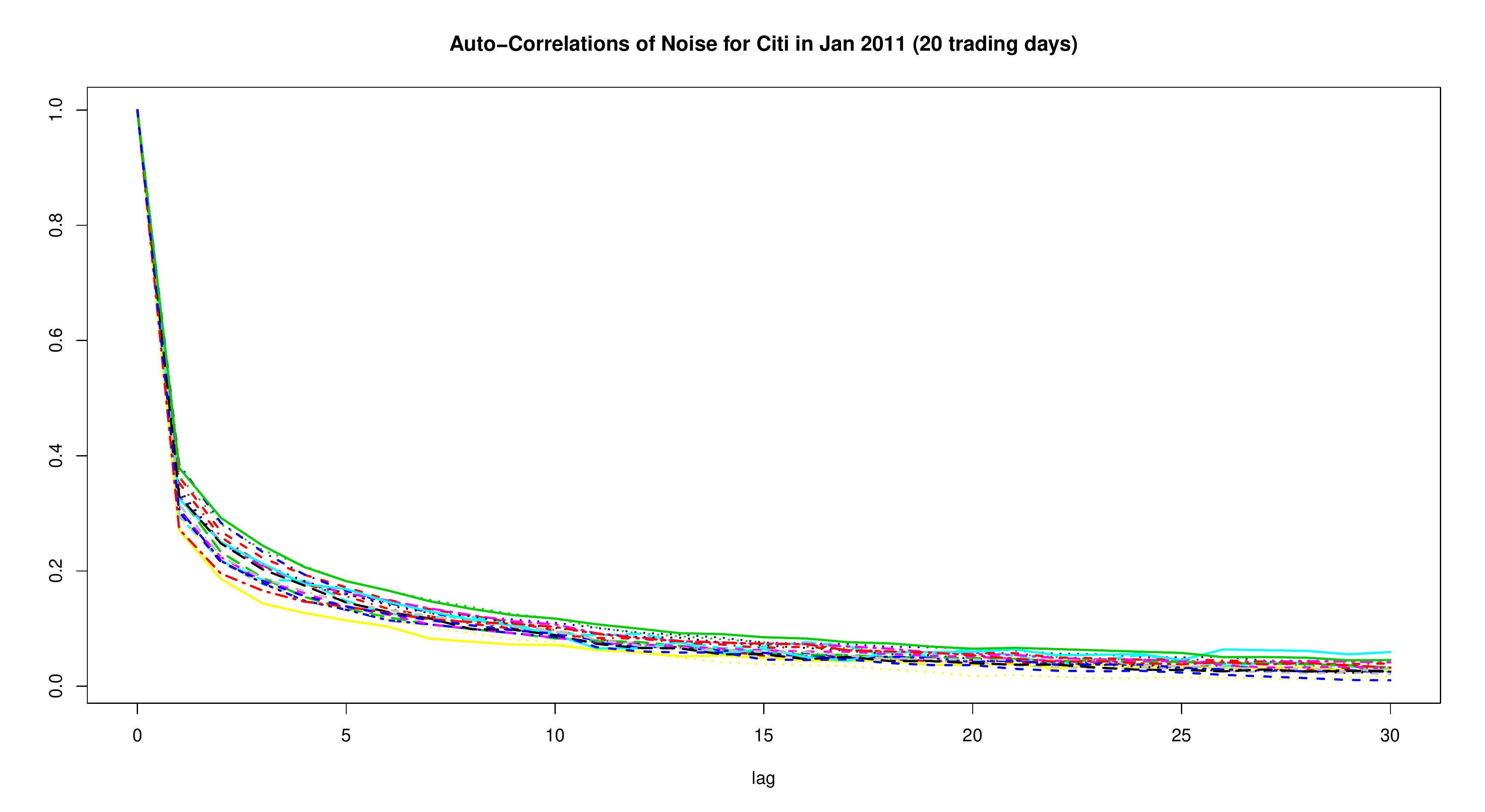}
\end{center}
  \caption{Estimates of auto-covariances and auto-correlations for the
  tick-by-tick trade data of Citigroup in Jan 2011. Each curve is for one
  trading day, and plots the estimates of auto-covariances (upper) or
  auto-correlations (lower)  of orders 0 through 30.}
\label{Fig:est_Citi}
\end{figure}

We first analyze the  tick-by-tick trade data of Citigroup Inc. (NYSE: C) in
Jan 2011. The average observation frequency is about 246,000 per day ($T=1$).
The observation times are irregular (not equidistant). We assume that the
Assumption (NO-1) is satisfied. We estimate both the auto-covariances and
auto-correlations  of orders 0 to 30,  using $\wwr(j)_n$ as in \eqref{MA1}
and $\wCo(j)_n$ as in (\ref{MA7})(with $k_n=8$), for each of the 20 trading
days and plot them in Figure  \ref{Fig:est_Citi}. Each curve in Figure
\ref{Fig:est_Citi} represents one day.

The auto-correlations appears to decay in an exponential way, so we can
assume that the mixing condition that we put in Assumption (NO-1) is
satisfied. We see from the results  that the noise is not un-autocorrelated;
in fact, positively autocorrelated for all the days under study, at least
for small lags.

Based on Theorem \ref{TM1} we can further test whether the auto-covariances
are equal to zero. More specifically, under the null hypothesis
\[
  H_0: r(j) = 0, \q\mbox{for } j\geq 1,
\]
with $\bj=(0,j)$, we have by Theorem \ref{TM1} (and \eqref{N11}) that
\[
  \sqrt{\frac{N_n(T)}{\wSi_T^{\bj,\bj,n}}}\,\frac1{N_n(T)}~U(\bj)^n_T
~\tol~\n(0,1).
\]
So we can compute the $p$-value for testing $H_0: R(j) = 0$ as $P\Big(|Z| >
\sqrt{\frac{N_n(T)}{\wSi_T^{\bj,\bj,n}}}\,\frac1{N_n(T)}~U(\bj)^n_T\Big)$,
where $Z\sim N(0,1)$. For this dataset, we take $k_n=8$ and $k_n'=4$ in
estimating $\wSi_T^{\bj,\bj,n}$. The $p$-values turn out to be all close to
0 (most are extremely small, the biggest one is about $0.035$), for all
orders up to 30 and for all the 20 trading days under consideration. In
particular, since all the estimated auto-covariances are positive, the
results also imply that if one conducts a one-sided test
\[
  H_0: r(j) \leq 0, \q\mbox{for } j\geq 1,
\]
then one rejects these hypotheses at 0.05 significance level for all orders
up to 30 for the data under study. We hence conclude that the
auto-covariances are statistically significantly different from 0, and
actually, statistically significantly bigger than 0, for all orders up
to 30 and for all the 20 trading days under consideration

Based on Theorem \ref{TA1} we can conduct tests for the auto-correlations.
The results turn out to be the same as above, namely, the auto-correlations
are statistically significantly different from 0, and actually,
statistically significantly bigger than 0, for all orders up to 30 and
for all the 20 trading days under consideration.

Theorems \ref{TM1} and \ref{TA1} also allow us to build confidence bands
for the auto-covariances and auto-correlations. More specifically, based
on Theorem \ref{TM1} we can build 95\% confidence bands for the
auto-covariances $r(j)$ as
\bee\label{CB_cov}
  \left[\wwr(j) \pm 1.96\cdot \sqrt{\frac{\wSi_T^{\bj,\bj,n}}{N_n(T)}}
  \right],\q j=0,1,2,\ldots.
\eee
And similarly Theorem \ref{TA1} yields 95\% confidence bands for the
auto-correlations $\Co(j)$ as
\bee\label{CB_corr}
  \left[\wCo(j)_n \pm 1.96\cdot \sqrt{\frac{\ws(j)_T^n}{N_n(T)}}\right],
  \q j=1,2,\ldots.
\eee
Applying these formulae to the data on January 3, 2011 we then get the
confidence bands for  the auto-covariances and auto-correlations of the
noise, which we plot in Figure \ref{Fig:CB_Citi} below.
\begin{figure}[H]
\begin{center}
  \includegraphics[width=7cm,height=5cm]{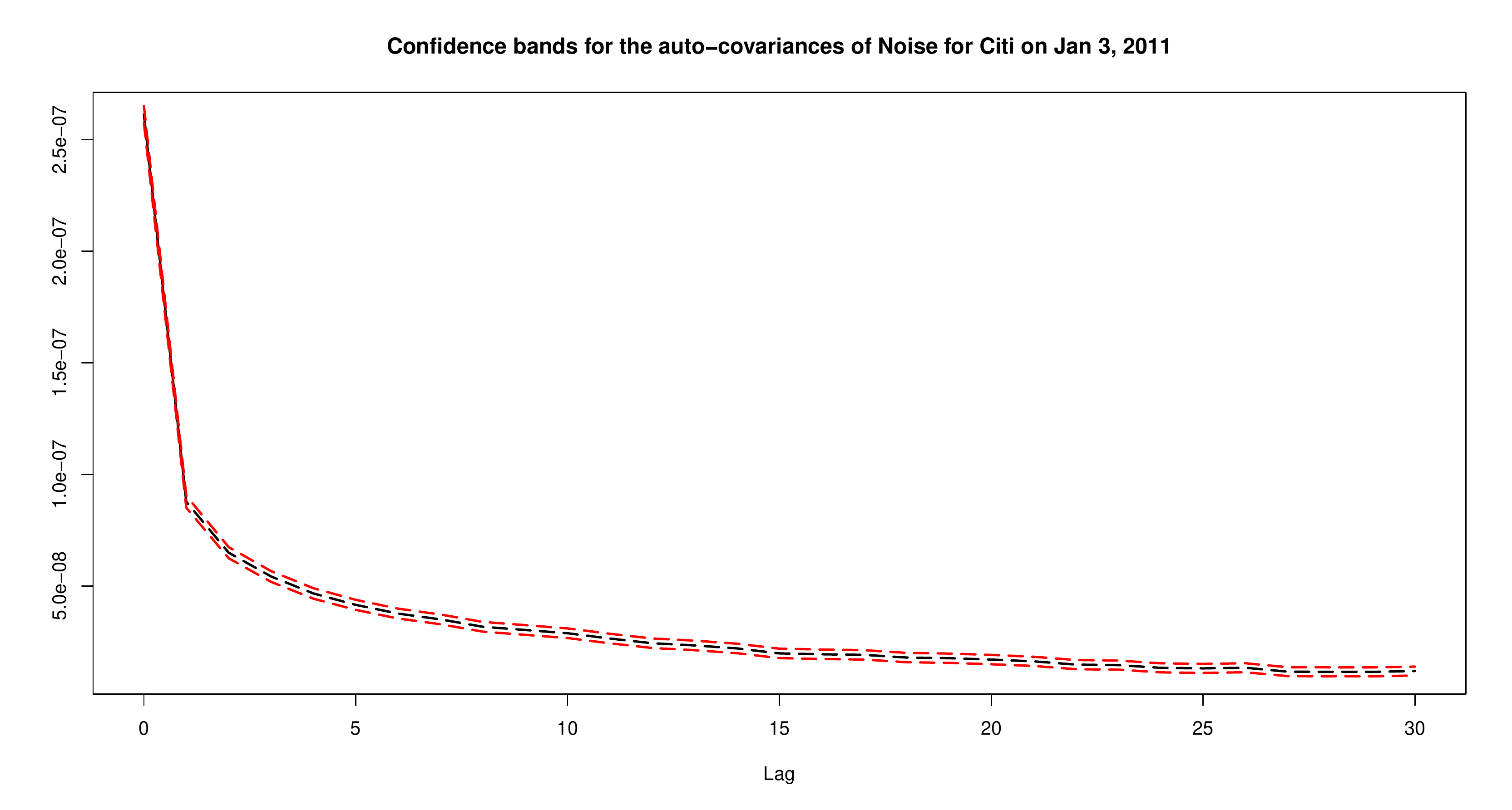}
  \includegraphics[width=7cm,height=5cm]{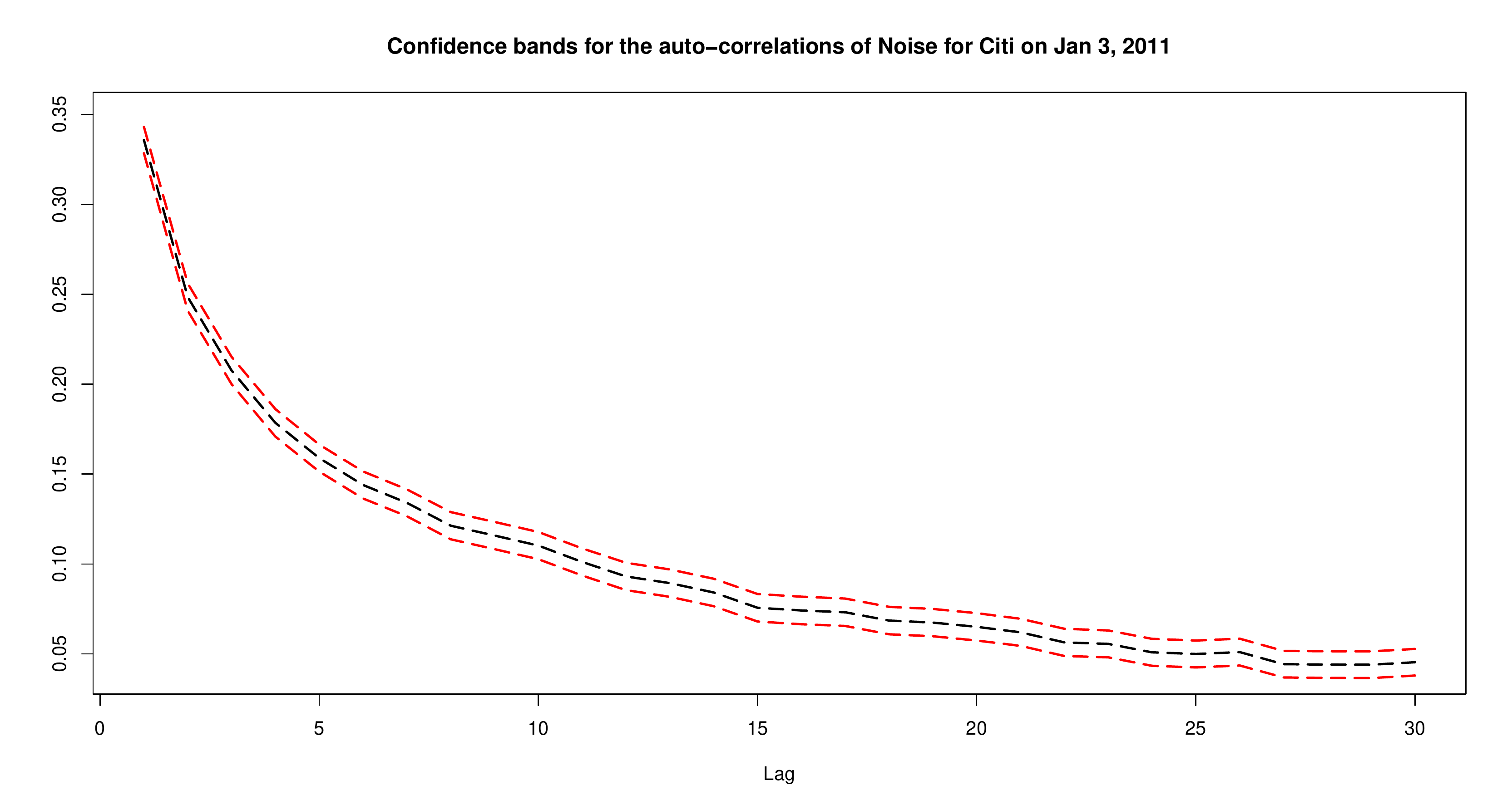}
\end{center}
  \caption{Confidence bands for the auto-covariances and auto-correlations
  for the tick-by-tick trade data of Citigroup on January 3, 2011. }
\label{Fig:CB_Citi}
\end{figure}
We can then conclude that, for example, on January 3, 2011 the
auto-correlation of the noise of any order up to 5 is greater than
0.15 or so, with 95\% confidence.

\subsection{Sprint-Nextel Jan 2011 Data}

\begin{figure}[H]
\begin{center}
  \includegraphics[height=6.2cm]{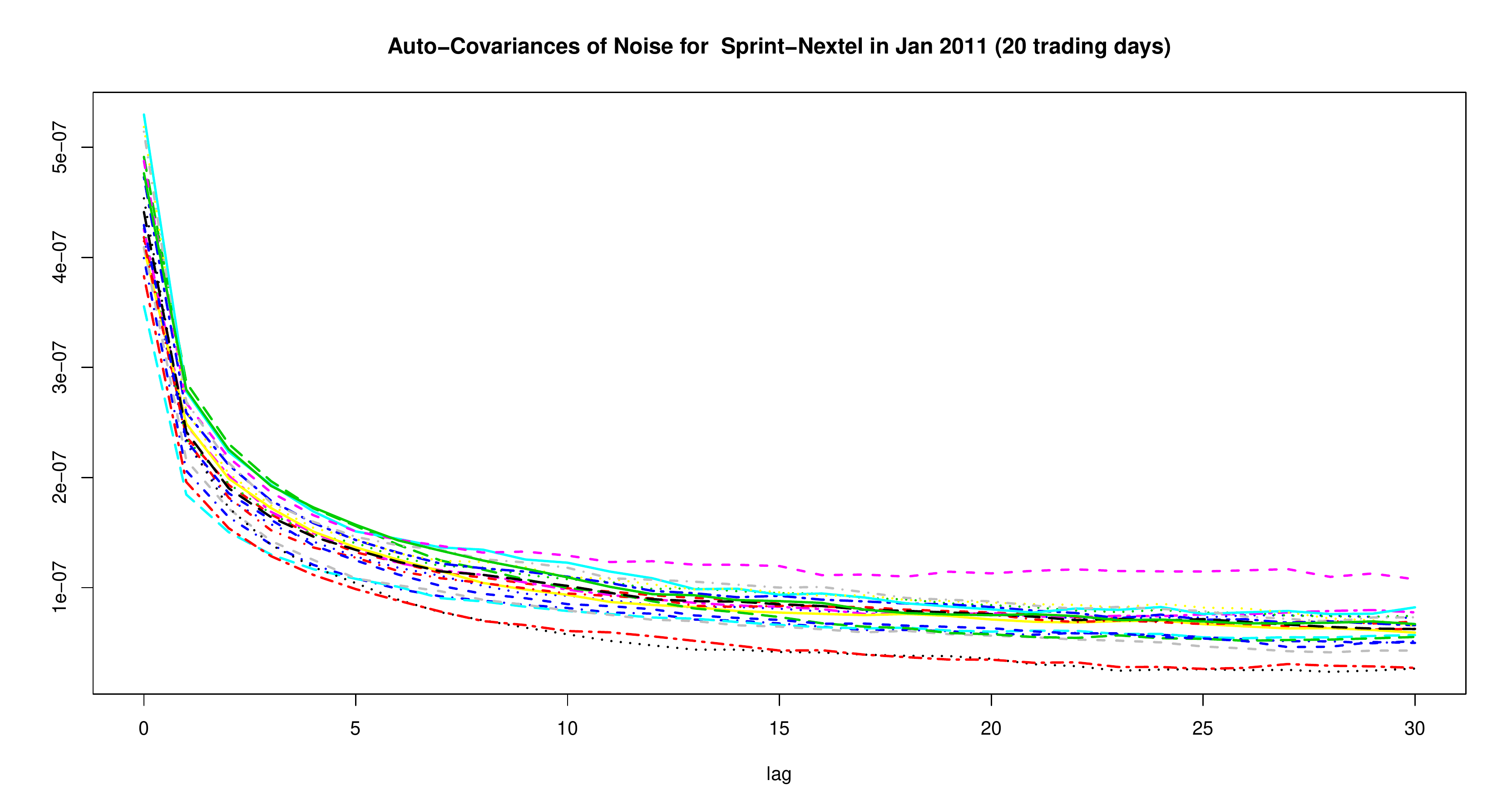}\\
  \includegraphics[height=6.2cm]{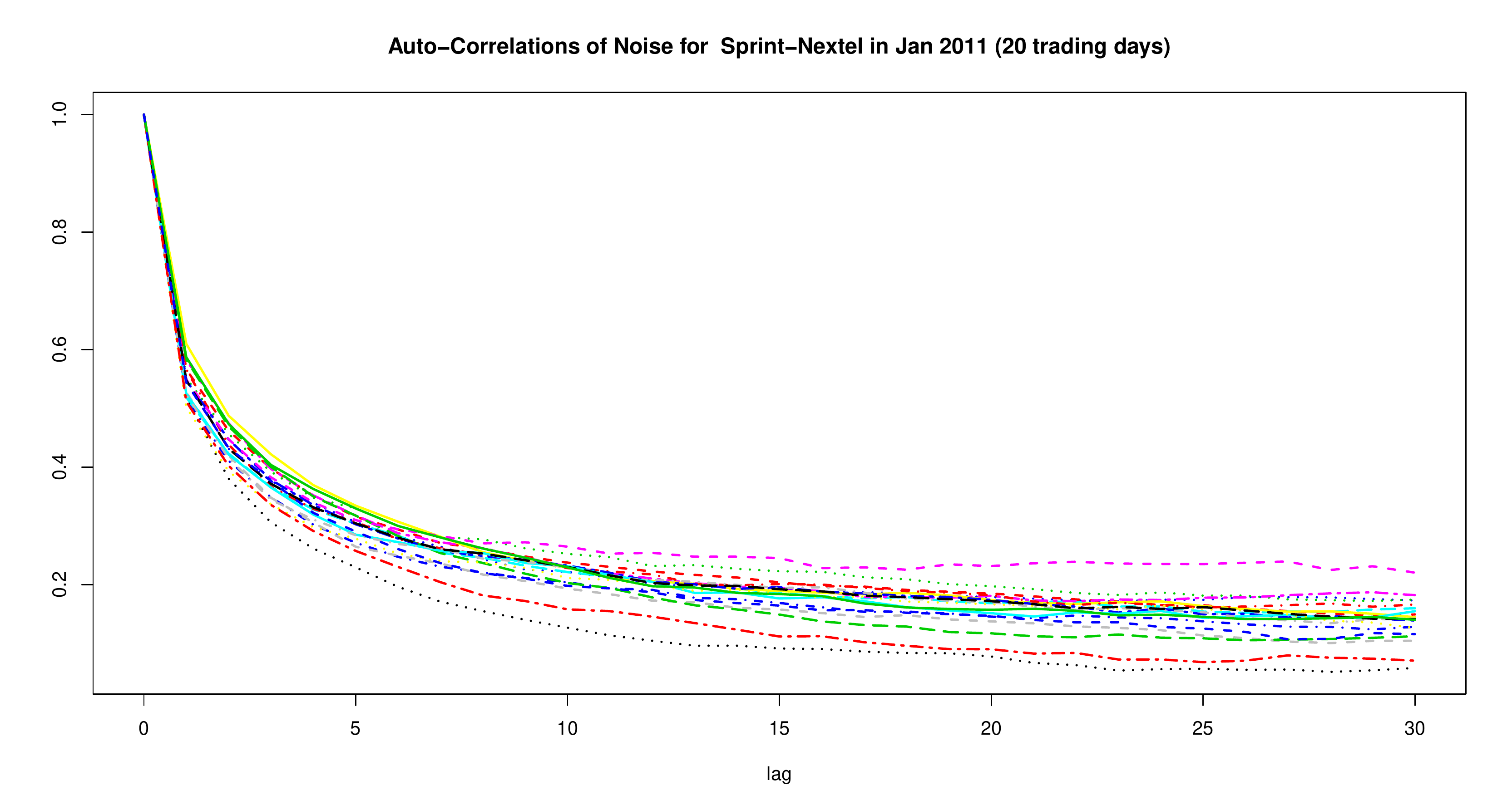}
\end{center}
  \caption{Estimates of auto-covariances (upper) and auto-correlations
  (lower) for the Sprint-Nextel trade data in Jan 2011. Each curve is for
  one trading day, and plots the estimates of auto-covariances or
  auto-correlations  of orders 0 through 30.}
\label{Fig:est_S}
\end{figure}

We next examine the tick-by-tick trade data of Sprint-Nextel Corporation
(NYSE:S) in Jan 2011. The average observation frequency is about 55,000 per
day ($T=1$). Assuming that the Assumption (NO-1) is satisfied,
we estimate both the auto-covariances and auto-correlations
of orders 0 to 30,  using $\wwr(j)_n$ as in \eqref{MA1} and $\wCo(j)_n$ as
in (\ref{MA7})(with $k_n=8$), for each of the 20 trading days and plot
them in Figure  \ref{Fig:est_S}. Again, each curve in Figure
\ref{Fig:est_S} represents one day.

We see similar phenomena as above, namely, (1) the auto-correlations decay
fairly quickly, and (2)  that the noise are not un-correlated; in fact,
positively correlated for all the days under study.

One can also conduct tests as in the previous subsection. The test results
are similar: for testing either the auto-covariances or auto-correlations
equal zero, the $p-$values are all extremely small (all smaller than
$10^{-5}$ in this case), and hence one can again conclude that the
auto-covariances/auto-correlations are statistically significantly
different from 0, and actually, statistically significantly bigger than 0,
for all orders up to 30 and for all the 20 trading days under consideration.

We can further construct confidence bands for the auto-covariances and
auto-correlations, using the formulae \eqref{CB_cov} and \eqref{CB_corr},
just as in the previous subsection. The resulting bands are plotted as
follows, again for the day of January 3, 2011:
\begin{figure}[H]
\begin{center}
  \includegraphics[width=7cm,height=5cm]{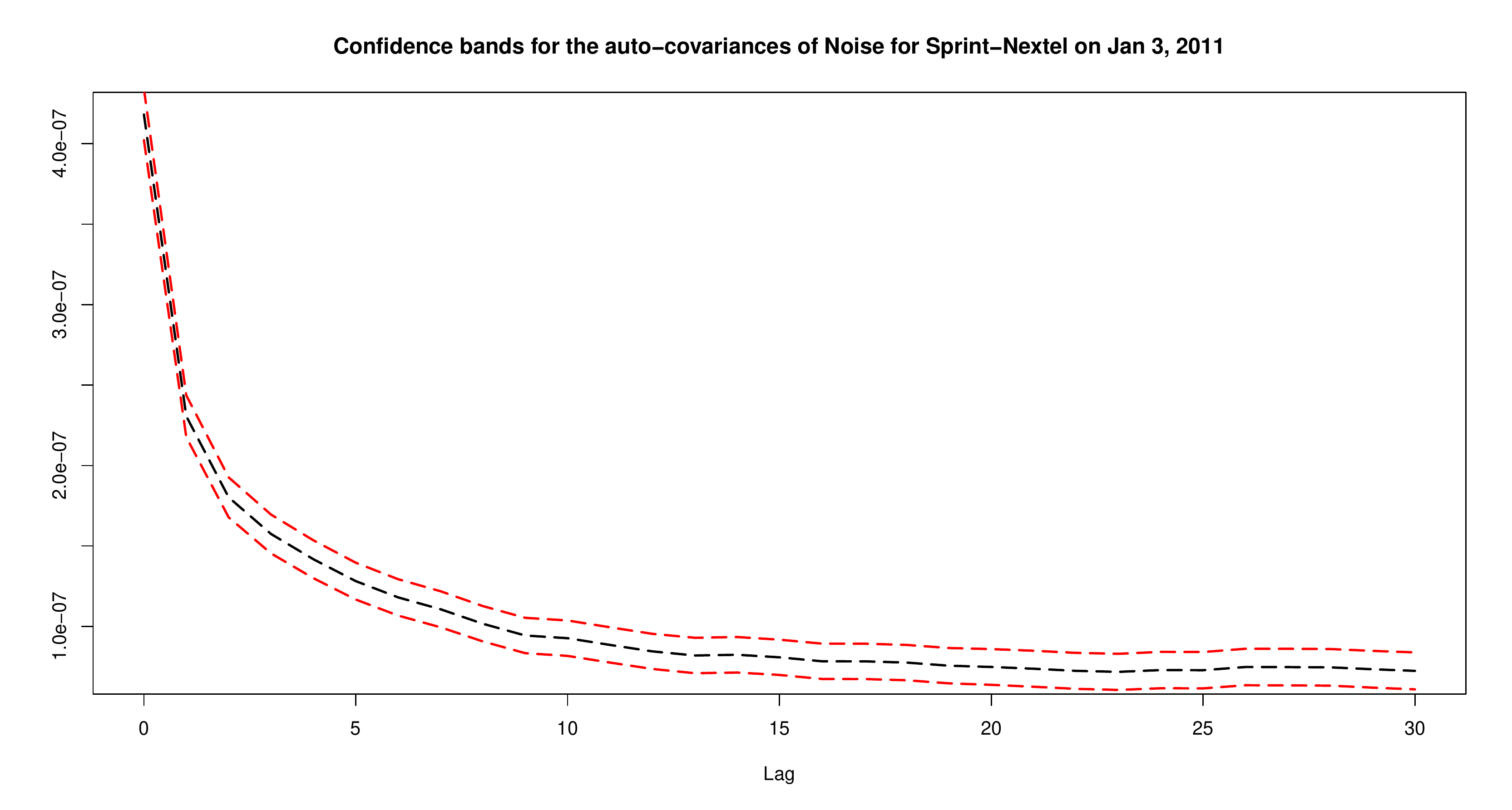}
  \includegraphics[width=7cm,height=5cm]{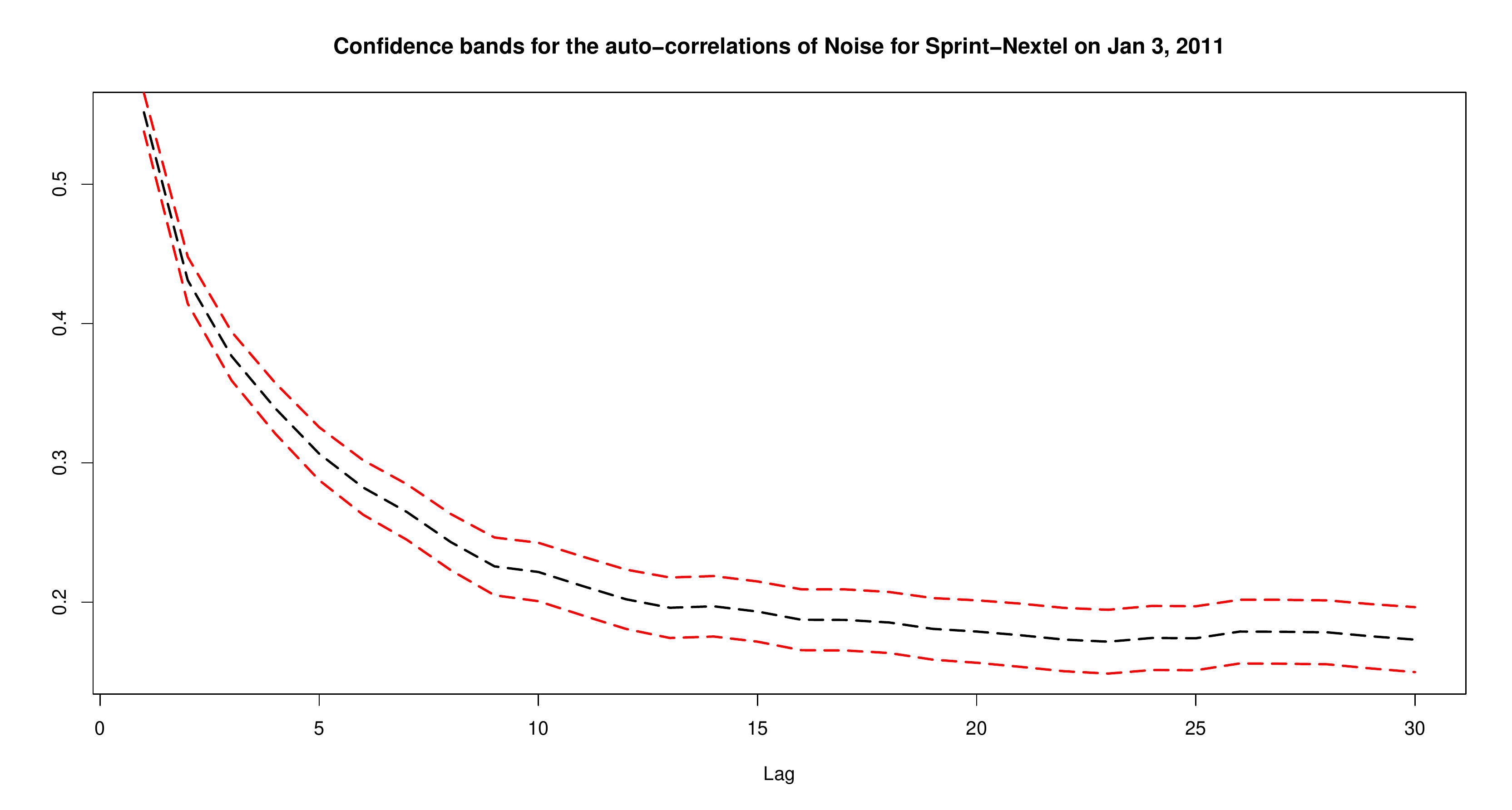}
\end{center}
  \caption{Confidence bands for the auto-covariances and auto-correlations
  for the tick-by-tick trade data of Sprint-Nextel on January 3, 2011. }
\label{Fig:CB_Sprint}
\end{figure}
Based on Figure \ref{Fig:CB_Sprint} we can conclude that on January 3, 2011
the auto-correlation of the noise of any order up to 5 is greater than 0.25
or so, with 95\% confidence.

\section{Conclusion and discussions}\label{sec:C}
In this paper we study the estimation of the (joint) moments, in particular,
the auto-covariances/auto-correlations of the microstructure noise, based
on high frequency data. We establish consistency as well as central-limit
theorems for our proposed estimators. Simulation studies demonstrate that
our estimators perform well. Empirical studies are also carried out,
in which by estimation and hypothesis testing that for the stocks tested,
the microstructure noises are not uncorrelated, but are actually (moderately)
positively correlated.

When the noises have general auto-correlations, the existing theory based
on i.i.d. noises or noises of other simple specific forms has to be
modified. In \cite{JLZ:vol}, the authors study how the noise structure
affects the estimation of volatility, and propose a volatility estimator
under Assumption (NO-2). Much more has to be done to better understand the
impact of the dependence structure of the market microstructure noise to
further financial applications.

\appendix

\renewcommand{\baselinestretch}{1.2}
\section{Proofs}\label{sec:P}
\setcounter{equation}{0}
\renewcommand{\theequation}{\thesection.\arabic{equation}}

Before starting the proof we mention that, by using a classical
localization procedure, for proving all the previously stated results,
we can replace the three assumptions (H), (N0-1) and (NO-2) by the
following stronger assumptions:
\vsc

\nib Assumption (SH) \rm We have (H), the processes $b$ and $\si$ are
bounded, and the stopping time $\tau_1$ is identically infinite
(so $|\de(\om,t,z)|\wedge1\leq J(z)$, where $J=J_1$).
\vsc

\nib Assumption (SNO-1) \rm We have (NO-1) and
$T(n,i)-T(n,i-1)\leq A\De_n$ and $\De_nN_n(t)\leq A_t$ for all $t$,
for some constants $A$ and $A_t$.
\vsc

\nib Assumption (SNO-2) \rm We have (NO-2) and
the process $\ga$ satisfies (SH) and is bounded.
\vsc

We always assume these strengthened assumptions below, mostly without
special mention. We also always assume, without mention, that
$(\chi_i)_{i\in\Z}$ is as in (NO-1). In all the sequel, the constant $K$
may vary from line to line, but does not depend on $n$ and on the
various indices $i,j,\cdots$.

Whether the noise and the underlying process $X$ are {\em a priori}\
defined on the same space or not is irrelevant for the results. However,
for the proofs it is convenient to suppose that $X$ and $\ga$ are defined
(and satisfy the relevant assumptions) on a space
$(\Om^{(0)},\f_\infty,(\f_t)_{t\geq0},\PP^{(0)})$,
whereas the sequence $(\chi_i)_{i\in\Z}$ is defined on another space
$(\Om^{(1)},\g,(\g_i)_{i\in\Z},\PP^{(1)})$, with
$\g_i=\si(\chi_k:j\leq i)$ and $\g^i=\si(\chi_k:j\geq i)$, and we set
$$\Om=\Om^{(0)}\times\Om^{(1)},\qquad
\f=\f_\infty\otimes\g,\qquad\PP=\PP^{(0)}\otimes\PP^{(1)}.$$
As usual, any variable or process {or} $\si$-field on $\Om^{(j)}$ for
$j=1,2$ is also considered, with the same notation, as defined on the
product $\Om$.

The space $(\Om^{(1)},\f^{(1)},(\g_i)_{i\in\Z},\PP^{(1)})$ is naturally
endowed with a measure-preserving and invertible transformation $\te$
such that $\chi_{i+j}=\chi_i\!\circ\te^j$ for all $i,j\in\Z$, and $\te$ is
ergodic by the
$\rho$-mixing property. Let us also recall a consequence of the definition
of the mixing coefficients~$\rho_j$, and of the product structure of the
space $(\Om,\f,\PP)$. We set $\h_i=\f_\infty\otimes\g_i$.
If $\xi$ is a {\em centered}\, square-integrable variable on
$(\Om^{(1)},\f^{(1)},(\g_i)_{i\in\Z},\PP^{(1)})$, which is measurable
with respect to $\g^0=\si(\chi_i:i\geq0)$, we have
\bee\label{P2}
\E\big(|\E(\xi\circ\te^{i+j}\mid\h_i)|^2\big)
~\leq~\rho_j^2\,\E(\xi^2)~\leq~\frac{K\,\E(\xi^2)}{j^{2v}}.
\eee
This yields the following useful estimate: if $\xi$ is $\g_i$-measurable
and $\xi'$ is $\g^{i+j}$-measurable, both square-integrable,
by (\ref{S0}) applied to $U=(\xi\circ\te^{-i}-E(\xi))
/\sqrt{\text{var}(\xi)}$ and $U'=(\xi'\circ\te^{-i}-E(\xi'))
/\sqrt{\text{var}(\xi')}$, plus $\rho_k(\chi)\leq K/k^v$ and
$\E(U)=\E(U')=0$ and by the stationarity, we have
\begin{align}
\big|\,\E(\xi\,\xi')\big|&=\big|\E(\xi)\,\E(\xi')
+\E(\xi)\,\E(U')\,\sqrt{\text{var}(\xi')}
+\E(\xi')\,\E(U)\,\sqrt{\text{var}(\xi)}\nonumber\\
&\qquad\qquad
+\E(UU')\,\sqrt{\text{var}(\xi)\,\text{var}(\xi')}\,\big|\nonumber\\
&\leq \big|\,\E(\xi)\,\E(\xi')\big|
+\frac K{j^v}\,\sqrt{\E(\xi^2)\,\E(\xi'^2)}.\label{P4}
\end{align}

Finally, recall that, if $V$ is a semimartingale
on $(\Om^{(0)},\f^{(0)},(\f_t)_{t\geq0},\PP^{(0)})$ which satisfies (SH),
then for any finite stopping time $S$ we have for $q\geq2$ and $t\geq0$:
\bee\label{P1}
\big|\,\E(V_{S+t}-V_S\mid\f_S)\big|\leq Kt,\qquad
\E\Big(\sup_{s\in[0,t]}|V_{S+s}-V_S|^q\mid\f_S\Big)\leq K_q\,t.
\eee

\subsection{Some Result about Stationary Processes.}

In this subsection we consider a sequence
$\xi^n=(\xi^{n,j})_{1\leq j\leq d}$ of $d$-dimensional variables on
the space $(\Om^{(1)},\f^{(1)},\PP^{(1)})$, satisfying the following, where
$w_n$ and $w$ are integers with
$w_n\geq w\geq~0$:
\bee\label{PS1}
\begin{array}{l}
\xi^n~\toop~\xi,\qquad\E(\xi^n)=0,\qquad
\sup_{n\in\N}\,\E(\|\xi^n\|^p)<\infty~~\forall\,p>0\\
\xi^n~~\text{is measurable with respect to the $\si$-field
$\g^0\cap\g_{w_n}=\si(\chi_i:0\leq i\leq w_n)$}\\
\xi~~\text{is measurable with respect to the $\si$-field
$\g^0\cap\g_{w}=\si(\chi_i:0\leq i\leq w)$}.
\end{array}
\eee
Note that $\E(\xi)=0$, whereas $w_n\to w$ is {\em not}\ assumed.
We write $\xi^n_i=\xi^n\circ\te^i$ and $\xi_i=\xi\circ\te^i$.
Then (\ref{P2}) and (\ref{P4}) yield
\bee\label{P3}
\begin{array}{l}
\E\big(\|\E(\xi^n_{m+i}\mid\h_m)\|^2\big)~=~
\E\big(\|\E(\xi^n_{m+i}\mid\g_m)\|^2\big)
~\leq~K\E(\|\xi^n\|^2)/i^{2v}~\leq~K/i^{2v}\\
\big|\,\E(\xi^{n,j}_m\,\xi^{n,k}_{m+i})\big|
~\leq~K\E(\|\xi^n\|^2)/((i-w_n) \vee 1)^v
~\leq~K/((i-w_n) \vee 1)^v
\end{array}
\eee
if $i\geq 1$, and the same for $\xi_{m+i}$ with $w_n$ replaced by $w$ in
the second inequality. Since $v>1$ we deduce that the
following define a covariance matrix:
\bee\label{PS2}
a^{jk}=\E(\xi_0^{j}\,\xi_0^{k})+\sum_{i=1}^\infty\big(\E(\xi_0^{j}\,\xi_i^{k})
+\E(\xi_i^{j}\,\xi_0^{k})\big).
\eee

In the simple situation where $\xi^n=\xi$ and $w_n=0$ (so $\xi$ is a
function of $\chi_0$), a trivial multi-dimensional
extension of Corollary VIII.3.106 of \cite{JS} yields a Central Limit
Theorem which says that
\bee\label{PS3}
\frac1{\sqrt{u_n}}\,\sum_{i=1}^{[tu_n]}\xi_i ~\toll~B,
\eee
for any sequence $u_n\to\infty$, where
\bee\label{PS4}
\text{$B$ a $d$-dimensional Brownian motion
with covariance}~~\E(B^j_1\,B^k_1)=a^{jk}.
\eee

We need to extend this result when $\xi^n$ depends on $n$, subject to
(\ref{PS1}), and $w_n\geq0$; this leads us to consider the following
processes, where as above $u_n$ is a sequence tending to $\infty$ and
$u'_n\geq0$ is another sequence of integers such that $u'_n/u_n\to0$:
\bee\label{PS51}
G^n=(G^{n,j})_{1\leq j\leq d},\quad\text{where}~~
G^{n,j}_t=\frac1{\sqrt{u_n}}\,
%%%\sum_{i=1}
\sum_{i=0}
^{[u_nt]-u'_n}\xi_i^{n,j}.
\eee
This will accommodate Theorem \ref{TM1}, whereas for Theorem \ref{TM2}
we additionally have random weights. In this case the observations are
equally spaced, and we need a normalization connected with the ``calendar''
time. So we set, with $u'_n\geq0$ a sequence
of integers such that $u'_n\De_n\to0$,
\bee\label{PS5}
H^n=(H^{n,j})_{1\leq j\leq d},\quad\text{where}~~
H^{n,j}_t=\rdn\,
%%%\sum_{i=1}
\sum_{i=0}
^{[t/\De_n]-u'_n} V^j_{i\De_n}\,\xi_i^{n,j},
\eee
where $V=(V^j)_{1\leq j\leq d}$ is a $d$-dimensional bounded
It\^o semimartingale satisfying (SH)
on $(\Om^{(0)},\f^{(0)},(\f_t),\PP^{(0)})$.

\begin{theo}\label{TPS1}
(a) Under (\ref{PS1}) and (NO-1), and if $u_n\to\infty$
and $w_n^2/u_n\to0$ and $u'_n/u_n\to0~$, for any
$t>0$ the variables $G^n_t$ converge in law to
$\n(0,at)$, with the matrix $a$ defined by \eqref{PS2}.

(b) Under (\ref{PS1}) and (NO-2), and if $w_n^2\De_n\to0$ and
$u'_n\De_n\to0$, for any $t>0$ the variables $H^n_t$ converge
$\f_\infty$-stably in law to a variable $H_t$ defined
on an extension $(\WOm,\Wf,\WP)$ of the space, and which conditionally on
$\f_\infty$ is centered Gaussian with (conditional)
covariance
\bee\label{PS6}
\WE(H^j_t\,H^k_t\mid\f_\infty)~=~a^{jk}\int_0^tV^j_s\,V^k_s\,ds.
\eee
\end{theo}

A way of realizing the limit $H$ above is to take $B$ as in (\ref{PS4})
and independent of $\f_\infty$, and to put
\bee\label{PS7}
H^j_t~=~\int_0^tV^j_s\,dB^j_s.
\eee
\vsd

\nib Proof. \rm 1) We start with (b), which is more complicated than (a).
By (\ref{P3}) and the Cauchy-Schwarz inequality, $\sum_{i\geq1}
\E(\|\E(\xi^n_i\mid\h_m)||)<\infty$ for any $m$, whereas $V$ is
bounded, so the following $d$-dimensional variables $U^n_m$ and $M^n_m$ are
well defined, componentwise:
$$\begin{array}{l}
U^{n,j}_m=\rdn
\sum_{i=(m-w_n)^+}^\infty
V^j_{i\De_n}\,\E(\xi^{n,j}_i\mid\h_m)\\
M^{n,j}_m=\rdn
\sum_{i=0}^\infty
V^j_{i\De_n}\big(\E(\xi_i^{n,j}\mid\h_m)-\E(\xi_i^{n,j}\mid\h_0)\big)
\end{array}$$
{and we write $\BM^n_m$ for the same variables as $M^n_m$, with $\xi^n$
substituted with $\xi$. Since $\xi^n_i$ is $\h_{i+w_n}$-measurable, we have
$\E(\xi_i^n\mid\h_{[t/\De_n]+w_n -u'_n+1})=\xi_i^n$ when
$i\leq[t/\De_n]-u'_n$, hence
\bee\label{PS9}\,
H^{n}_t=
M^{n}_{[t/\De_n]+w_n-u'_n+1}+U^{n}_0
-U^{n}_{[t/\De_n]+w_n-u'_n+1}.
\eee
\vsd

2) In this step we prove that, for any $t>0$, we have
\bee\label{PS1200}
U^n_{[t/\De_n]+w_n-u'_n+1}~\toop~0,
\q \mbox{and}\q U^n_{0}~\toop~0.
\eee
We shall only prove the first convergence as the second can be proved
similarly. To this end, we write $m_n=[t/\De_n]+w_n-u'_n+1$, and
since $u'_n\De_n\to0$ and $t>0$ we may assume $m_n>w_n$. Then we have
$U^n_{m_n}=A_n+B_n$, where
$$A_n=\rdn\,\sum_{i=1+m_n}^\infty V^j_{i\De_n}\,
\E(\xi^{n,j}_i\mid\h_{m_n}),\qquad
B_n=\rdn\,\sum_{i=m_n-w_n}^{m_n}
V^j_{i\De_n}\,\E(\xi^{n,j}_i\mid\h_{m_n}).$$

Since $V$ is bounded, we deduce from \eqref{P3} and Cauchy-Schwarz
inequality that
$$\E(\|A_n\|^2)\leq K\De_n\sum_{i,j=1+m_n}^\infty
\E\big(\|\E(\xi^{n}_i\mid\h_{m_n})\|\,
\|\E(\xi^{n}_j\mid\h_{m_n})\|\big)\leq K\De_n,$$
hence $A_n\toop0$. Next, we have $\E(\|\xi^n_i\|^2)\leq K$ by \eqref{PS1},
hence $\E(\|B_n\|^2)$ is obviously smaller than $K\De_nw_n^2$ because the
sum defining $B_n$ contains $w_n$ terms. Since $\De_nw_n^2\to0$ we deduce
$B_n\toop0$, hence the first convergence in \eqref{PS1200}.
\vst

3) In this step
%%%, and with $m_n$ as above,
we prove
\bee\label{PS12}
\|M^n_{[t/\De_n]+w_n-u'_n+1}-\BM^n_{[t/\De_n]+w_n-u'_n+1}\|
~\toop~0.
\eee
Setting $\xi'^n_i=\xi^n_i-\xi_i$ and again $m_n=[t/\De_n]+w_n-u'_n+1$,
we observe that $M^n_{m_n}-\BM^n_{m_n}=\sum_{k=1}^{m_n}\eta^n_k$, where
$$\eta^{n,j}_k~=~\rdn\,
\sum_{i\geq 0}
V^j_{i\De_n}\,
\big(\E(\xi'^{n,j}_i\mid\h_k)-\E(\xi'^{n,j}_i\mid\h_{k-1})\big)$$
is a martingale increment, relative to the discrete time filtration
$(\h_k)_{k\geq0}$, hence $\E((M^{n,j}_{m_n}-\BM^{n,j}_{m_n})^2)=
\sum_{k=1}^{m_n}\E((\eta^{n,j}_k)^2)$. By successive conditioning,
$$\begin{array}{l}
\sum_{k=1}^{m_n}\E((\eta^{n,j}_k)^2)=
\De_n\E\Big(\sum_{k=1}^{m_n}
\sum_{i,l\geq 0}
V^j_{i\De_n}\,V^j_{l\De_n}
\,\big(\E(\xi'^{n,j}_i\mid\h_k)\,\E(\xi'^{n,j}_l\mid\h_k)\\
\hskip7cm
-\E(\xi'^{n,j}_i\mid\h_{k-1})\,\E(\xi'^{n,j}_l\mid\h_{k-1})\big)\Big)
\end{array}$$
and the double series $\sum_{i,l}$ inside the expectation
above is absolutely convergent (almost surely). Hence we
may permute the order of summation over $(i,l)$ and over $k$, and thus
get $\E((M^{n,j}_{m_n}-\BM^{n,j}_{m_n})^2)=D^n_{m_n}-D^n_0$, where
$$D^n_k=\De_n\E\Big(
\sum_{i,l\geq 0}
V^j_{i\De_n}\,V^j_{l\De_n}
\,\E(\xi'^{n,j}_i\mid\h_{k})
\,\E(\xi'^{n,j}_l\mid\h_{k}\Big).$$

We have $D^n_k=\sum_{r=1}^7D(r)^n_k$, where (with an empty sum set to $0$)
$$\begin{array}{l}
D(1)^n_k=\De_n\E\Big(\sum_{i\geq k+1}(V^j_{i\De_n})^2
\,(\E(\xi'^{n,j}_i\mid\h_{k}))^2\Big)\\
D(2)^n_k=\De_n\E\Big(\sum_{0\vee(k-w_n)<i\leq k}(V^j_{i\De_n})^2
\,(\E(\xi'^{n,j}_i\mid\h_{k}))^2\Big)\\
D(3)^n_k=\De_n\E\Big(
\sum_{0\leq i\leq k-w_n}
(V^j_{i\De_n})^2
\,(\E(\xi'^{n,j}_i\mid\h_{k}))^2\Big)\\
D(4)^n_k=2\De_n\E\Big(\sum_{l>i>k} V^j_{i\De_n}\,V^j_{l\De_n}
\,\E(\xi'^{n,j}_i\mid\h_{k})\,\E(\xi'^{n,j}_l\mid\h_{k})\Big)\\
D(5)^n_k=2\De_n\E\Big(\sum_{0\vee(k-w_n)<i\leq k}\sum_{l>k}
V^j_{i\De_n}\,V^j_{l\De_n}
\,\E(\xi'^{n,j}_i\mid\h_{k})\,\E(\xi'^{n,j}_l\mid\h_{k})\Big)\\
D(6)^n_k=2\De_n\E\Big(\sum_{0\vee(k-w_n)<i<l\leq k}
V^j_{i\De_n}\,V^j_{l\De_n}
\,\E(\xi'^{n,j}_i\mid\h_{k})\,\E(\xi'^{n,j}_l\mid\h_{k})\Big)\\
D(7)^n_k=2\De_n\E\Big(
\sum_{0\leq i\leq k-w_n}
\sum_{l>i}
V^j_{i\De_n}\,V^j_{l\De_n}
\,\E(\xi'^{n,j}_i\mid\h_{k})\,\E(\xi'^{n,j}_l\mid\h_{k})\Big).
\end{array}$$
Since $V$ is bounded, \eqref{P3} applied with $\xi'^n$ and with
$\al_n=\E(\|\xi^n-\xi\|^2)$ yields $D(r)^n_k\leq K\De_n \al_n$
for $r=1,4$ (recall $v>1$). Next, one has
$D(r)^n_k\leq K\De_n \al_nw_n$ for $r=2$
and also for $r=5$ by applying \eqref{P3} again, whereas
$D(6)^n_k\leq K\De_n\al_nw_n^2$. Finally, since $\E(\xi'^n_i\mid\h_k)=
\xi'^n_i$ when $i\leq k-w_n$, one has $D(3)^n_k\leq K\De_n \al_n(1+
(k-w_n)^+)$ and one can rewrite $D(7)^n_k$ as
$$D(7)^n_k=2\De_n\E\Big(\sum_{0\leq i\leq k-w_n}
\sum_{l>i}V^j_{i\De_n}\,V^j_{l\De_n}
\,\xi'^{n,j}_i\,\E(\xi'^{n,j}_l\mid\h_{i+1})\Big),$$
and another application of \eqref{P3}  yield
$D(7)^n_k\leq K\De_n\al_n (1+(k-w_n)^+)$.
Putting all these estimates together (for $k=0$ and
for $k=m_n$, and since $w_n\geq1$) gives us
$$\E((M^{n,j}_{m_n}-\BM^{n,j}_{m_n})^2)\leq K \De_n\al_n
\big(w_n^2+[t/\De_n]\big).$$
We have $\al_n\to0$ and $\De_nw_n^2\to0$ by hypothesis, and
\eqref{PS12} follows.
\vst

4) In view of (\ref{PS9}), \eqref{PS1200} and (\ref{PS12}) it remains to
prove the $\f_\infty$-stable convergence of the variables
$\BM^n_{[t/\De_n]+w_n-u'_n+1}$, and we actually prove a stronger result.
Namely, we will show the $\f_\infty$-stable convergence of the processes
$\BM^n_{[t/\De_n]}$ to a process $H$ which conditionally on $\f_\infty$
is a centered continuous Gaussian martingale with covariance given by
\eqref{PS6} for any $t\geq0$ (since $\De_n(w_n-u'_n+1)\to 0$
this implies the convergence of
$\BM^n_{[t/\De_n]+w_n-u'_n+1}$ toward $H_t$).

As in Step 3, $\BM^n_l=\sum_{m=1}^l\ze^n_m$, where
$$\ze_m^{n,j}=\rdn\,
\sum_{i=0}^\infty
V^j_{i\De_n}\,\be^{j}_{i,m},
\quad \be^{j}_{i,m}=
\E(\xi_i^{j}\mid\h_m)-\E(\xi_i^{j}\mid\h_{m-1}),$$
and each $\ze^n_m$ is a martingale increment. Hence, if
$$c_m^{n,jk}=\E(\ze^{n,j}_m\,\ze^{n,k}_m\mid\h_{m-1}),\qquad
c(\ep)_m^n=\E(\|\ze^{n}_m\|^2\,1_{\{\|\ze^n_m\|>\ep\}}\mid\h_{m-1}),$$
we deduce from Theorems VIII.3.22 and VIII.5.14 of \cite{JS}
the $\f_\infty$-stable convergence of the processes
$\BM^n_{[t/\De_n]}$ to $H$, as soon as we have the
following two properties, for all $t,\ep>0$:
\bee\label{PS14}
\sum_{m=1}^{[t/\De_n]} c_m^{n,jk}~\toop~a^{jk}\int_0^tV^j_s\,V^k_s\,ds\,,
\qquad \sum_{m=1}^{[t/\De_n]}c(\ep)_m^n~\toop~0.
\eee

The second one is easy to prove. Indeed, if
$\Wbe_m=\sum_{i\in\Z}\|\be_{i,m}\|$, we have $\|\ze^n_m\|\leq A\rdn\,\Wbe_m$
for some $A>0$: we allow the index $i$ to be
negative, so that we can apply the obvious relation
$\be_{i,m}=\be_{i-m,0}\circ\te^m$ (for all $i,m\in\Z$) to obtain
$\Wbe_m=\Wbe_0\circ\te^m$. {\em A priori}\ $\Wbe_m$ could be infinite,
however $\be_{i,m}=0$ when $i< m-w$ by (\ref{PS1}), so (\ref{P3}) for
$\xi$ implies that $\E((\Wbe_m)^2)\leq K$. Then by stationarity
$$\begin{array}{c}
\E(c(\ep)^n_m)\leq A^2\De_n\E\big(\Wbe_m^2
\,1_{\{\Wbe_m>\ep/(A\sqrt{\De_n})\}}\big)=A^2\De_n\Bal(\ep)_n\\
\text{where}~~\Bal(\ep)_n=\E\big((\Wbe_0)^2
\,1_{\{\Wbe_0>\ep/(A\sqrt{\De_n})\}}\big).
\end{array}$$
Now, $\Bal(\ep)_n\to0$ because $\E(\Wbe_0^2)\leq K$, and the second
part of (\ref{PS14}) follows.
\vst

5) By virtue of the square-integrability of $\Wbe_m$, the $d$-dimensional
variables $\Bbe_m=\sum_{i\geq 0}\be_{i,m}$ are well-defined, square-integrable,
and also $\Bbe_m=\Bbe_{w+1}\circ\te^{m-w-1}$ for all $m\geq w+1$ (this
is wrong when $1\leq m\leq w$). In this step, we show that
\bee\label{PS125}
|\E(\Bbe_m^j\,\Bbe_m^k)|\leq K,\qquad
\mbox{and if }~
m\geq w+1,
\mbox{ then }~
\E(\Bbe_m^j\,\Bbe_m^k)=a^{jk},
\eee
with $a^{jk}$ given by (\ref{PS2}). The first estimate follows from
$\|\Bbe_m\|\leq\Wbe_m=\Wbe_0\circ\te^m$ and $\Wbe_0\in\LL^2$. For the
second property, by polarization it is enough to show it in
the one-dimensional case $d=1$, and so below we omit $j,k$. The variable
$\be_{i,m}=\E(\xi_i\mid\h_m)-\E(\xi_i\mid\h_{m-1})$ is $\h_m$-measurable
with vanishing $\h_{m-1}$-conditional mean, whereas
$\xi_{i+1}\,\E(\xi_{l+1}\mid\h_m)=\big(\xi_i\,\E(\xi_l\mid\h_{m-1})\big)
\circ\te$. Then
$$\E(\be_{i,m}\,\be_{l,m})=\E(\xi_i\,\be_{l,m})=
\E\big(\E(\xi_i-\xi_{i+1}\mid\h_m)\,\E(\xi_l\mid\h_m)\big)
+\E\big(\E(\xi_{i+1}\mid\h_m)\,\E(\xi_l-\xi_{l+1}\mid\h_m)\big),$$
hence for any $L>2m$:
$$\sum_{i,l=0}^L
\E(\be_{i,m}\,\be_{l,m})=
\sum_{l=0}^L \E\big(\E(\xi_0\mid\h_m)\,\E(\xi_l+\xi_{l+1}\mid\h_m)\big)
-\sum_{l=0}^L\E\big(\E(\xi_{L+1}\mid\h_m)\,\E(\xi_l+\xi_{l+1}\mid\h_m)\big).$$
By (\ref{P3}) the $l$th summand in the last sum above is smaller
in absolute value than $K/L^v$ always, and than $K/L^vl^v$ when
$l>2m$. Since $v>1$, by letting $L\to\infty$ we obtain that
$$\E((\Bbe_m)^2)=\sum_{l=0}^\infty
\E\big(\E(\xi_0\mid\h_m)\,\E(\xi_l+\xi_{l+1}\mid\h_m)\big)=
\E(\xi_0^2)+2\sum_{l=1}^\infty\E\big(\xi_0\,\xi_l\big),$$
the last equality following from the fact that $m\geq w+1$, hence $\xi_0$ is
$\h_m$-measurable. The right side above is (\ref{PS2}) in the one-dimensional
case, and thus the last part of (\ref{PS125}) holds.
\vsd

6) In this step we set $c'^{jk}_m=\E\big(\Bbe^j_m\,\Bbe^k_m\mid\h_{m-1}\big)$
and prove that
\bee\label{PS15}
\sum_{m=1}^{[t/\De_n]}\big(c_m^{n,jk}-
\De_n\,V_{(m-1)\De_n}^j\,V_{(m-1)\De_n}^k
\,c'^{jk}_m\big)~\toucp~0.
\eee
Letting $\eta^n_m$ be the $m$th summand above, we see that
$\eta^n_m=\De_n\sum_{i,l\geq 0}
\eta(i,l)^n_m$, where
$$\eta(i,l)^n_m=\big(V^j_{i\De_n}\,V^k_{l\De_n}-
V^j_{(m-1)\De_n}\,V^k_{(m-1)\De_n}\big)\,
\E\big(\be_{i,m}^{j}\,\be_{l,m}^{k}\mid\h_{m-1}\big).$$
As seen before, $\be_{i,m}=0$ when $i<m-w$ and $\E(\|\be_{i,m}\|^2)$ is
smaller than $K$ always, and than $K/(i-m)^{2v}$ when $i>m$; hence by
(\ref{P1}) and (\ref{P3}) we obtain if $i\leq l$
$$\E(|\eta(i,l)^n_m|)\leq \left\{\begin{array}{ll}
0&\text{if}~~i< m-w\\
K\frac{1\wedge\sqrt{\De_n((l-m)\vee|m-1-i|)}}{(l-m)^v}
&\text{if}~~m-w\leq i\leq m<l\\
K\frac{1\wedge\sqrt{\De_n(l-m)}}{(i-m)^v(l-m)^v}
&\text{if}~~i>m\\
K\rdn&\text{if}~~m-w\leq i\leq l\leq m,
\end{array}\right.$$
and similar estimates hold when $l\leq i$. Since one can always
assume $v\in(1,3/2)$, in which case
$\sum_{i\geq1}(1\wedge\sqrt{i\De_n}\,)/i^v\leq K\De_n^{v-1}$, we get
$\E(|\eta^n_m|)\leq K\De_n^{v}$, and (\ref{PS15}) follows.
\vsd

7) By the previous step, in order to get the first part of (\ref{PS14})
we are left to show
\bee\label{PS16}
\De_n\sum_{m=1}^{[t/\De_n]} V_{(m-1)\De_n}^j\,V_{(m-1)\De_n}^k\,c_m'^{jk}
~\toop~a^{jk}\int_0^tV^j_s\,V^k_s\,ds.
\eee
The left side above can be considered as the integral of the
c\`adl\`ag function $s\mapsto V^j_s\,V^k_s$ with respect to the (random)
measure $F^{n,jk}_t(ds)=\De_n\sum_{m=1}^{[t/\De_n]} c'^{jk}_m\textbf\,
\delta_{(m-1)\De_n}(ds)$,
where $\delta_{x}$ stands for the delta measure at $x$,
so it is enough to show that $F^{n,jk}_t$ converges in probability
to the measure $a^{jk}\,1_{[0,t]}(s)\,ds$. To this aim, it is is enough to
show that
\bee\label{PS17}
s\leq t~~\Rightarrow~~
G^n_s~:=~\De_n\sum_{m=1}^{[s/\De_n]}c'^{jk}_m~\toop~a^{jk}\,s
\eee
(this is obvious when $k=j$, because then $F^{n,jk}_t$
is a positive measure; when $k\neq j$ it may be a signed measure, but
with an absolute value dominated by $\frac12\,(F^{n,jj}_t+F^{n,kk}_t)$,
so again (\ref{PS17}) is enough).

We recall that $\Bbe_m=\Bbe_{w+1}\circ\te^{m-w-1}$ when $m>w$, implying
$c'_m=c'_{w+1}\circ\te^{m-w+1}$, whereas it is
obviously enough to show the convergence (\ref{PS17}) when the sum
starts at $m=w+1$. Then, the ergodic theorem and (\ref{PS125}) tell us that
$G^n_s$ converges a.s.
(locally uniformly in $s$) to  $a^{jk}s$. This completes the proof of (b).
\vsd

8) Now we turn to (a). This is basically the same as (b), with the
processes $V^j$ being identically equal to $1$, and with the convention
$\De_n=1/u_n$ (indeed, in this case, the calendar time and the observation
times $T(n,i)$ play no role at all, and neither does $\f_\infty$; so
(\ref{S2}) is irrelevant, and we
can set $\De_n=1/u_n$). So all Steps 1--7 can be reproduced, except Step 6
which is irrelevant, whereas in Step 7 we can proceed directly to
(\ref{PS17}).\qed
\vsc

We will also need bounds for the moments of the processes $G^n$ and $H^n$:

\begin{lem}\label{LPS1} Under (\ref{PS1}) and if $V$ is bounded, we have
$$\begin{array}{lll}
\text{\rm(NO-1)}\quad&\Rightarrow\quad&\E(\|G^n_t\|^2)~\leq~K(1+w_n)t\\
\text{\rm(NO-2)}\quad&\Rightarrow\quad&\E(\|H^n_t\|^2)~\leq~K(1+w_n)t.
\end{array}$$
\end{lem}

\nib Proof. \rm Upon setting $V^j_t=1$ and $u_n=1/\De_n$, the case (NO-1)
reduces to the case (NO-2). By singling out each component
$H^{n,j}$ we can assume $d=1$. Then we have
$\E(|H^n_t|^2)=\phi_n(t)+\psi_n(t)$, where
$$\begin{array}{l}
\phi_n(t)=\De_n\sum_{i=0}^{[t/\De_n]-u'_n}
\E\big(|V_{i\De_n}|^2\,|\xi^{n}_i|^2\big)
+2\De_n\sum_{0\leq i<l\leq(i+w_n)\wedge([t/\De_n]-u'_n)}
\E\big(V_{i\De_n}\,\xi^{n}_i\,V_{l\De_n}\,\xi^{n}_l\big)\\
\psi_n(t)=2\De_n\sum_{0\leq i<i+w_n<l\leq[t/\De_n]-u'_n}
\E\big(V_{i\De_n}\,\xi^{n}_i\,V_{l\De_n}\,\xi^{n}_l\big).
\end{array}$$

On the one hand, since $V$ is bounded and (\ref{PS1}) holds,
$|\phi_n(t)|\leq K(1+w_n)t$ follows from the Cauchy-Schwarz inequality. On the
other hand, $\xi^n_i$ is $\g_{i+w_n}$-measurable, so by conditioning
first with respect to $\f_\infty$ we deduce from (\ref{P3})
that $\big|\E\big(V_{i\De_n}\,\xi^{n}_i\,V_{l\De_n}\,\xi^{n}_l)\big|\leq K/(l-i-w_n)^v$ when $l>i+w_n$.
Since $v>1$ one deduces $\psi_n(t)\leq
Kt$, and the result follows.\qed
\vsc

Finally, we need to consider processes that are slightly more general
than $H^n$, at least in the one-dimensional case. Namely, we
assume (\ref{PS1}) with $d=1$, and we are also given another set
$(\xi'^n,w'_n)$ satisfying (\ref{PS1}) as well (we do not need a limit
$\xi'$ here), plus an
arbitrary sequence of integers $\rho_n\geq1$. With the same auxiliary
bounded process $V$ and sequence of integers $u'_n$ as above, we set
\bee\label{PS20}
\BH^n_t=\rdn\,\sum_{i=0}^{[t/\De_n]-u'_n}
 V_{i\De_n}\,\xi^n_i\,\xi'^n_{i+w_n+\rho_n}.
\eee

\begin{lem}\label{LPS2} In the above setting, and under (NO-2), we have
$$\E(|\BH^n_t|^2)~\leq~K\Big(1+w_n'+\frac{w_n}{\rho_n^v}+
\frac t{\De_n\,\rho_n^{2v}}\Big)t.$$
\end{lem}

\nib Proof. \rm Set $\xi''^n_i=\xi^n_i\,\xi'^n_{i+w_n+\rho_n}$.
By (\ref{PS1}) and (\ref{P4}) applied repeatedly, we check that
for $l\geq0$,
$$\big|\,\E(\xi''^n_i\,\xi''^n_{i+l}\mid\f_\infty)\big|\leq
\left\{\begin{array}{ll}
K&\text{if}~l\leq w'_n\\
K/(l-w'_n)^v&\text{if}~w'_n<l\leq w'_n+\rho_n\\
K/\rho_n^v&\text{if}~w'_n+\rho_n<l\leq w'_n+\rho_n+w_n\\
K(1/\rho_n^{2v}+1/(l-w'_n-w_n-\rho_n)^v)&\text{if}~l>w'_n+\rho_n+w_n.
\end{array}\right.$$
Since $V$ is bounded, we have
$$\E(|\BH^n_t|^2)\leq K\De_n\sum_{i=0}^{[t/\De_n]}~\sum_{l=0}^{[t/\De_n]}
\E\big(\big|\,\E(\xi''^n_i\,\xi''^n_{i+l}\mid\f_\infty)\big|\big).$$
Then by splitting the sum over $l$ according to the four cases described
above, we obtain the result.\qed

\subsection{{Further} Auxiliary Results}

In this subsection we gather a few results of a technical character, to
be used at several places.
\vst

\nib 1) \rm The first of these results is about asymptotically negligible
triangular arrays. The setting is as follows: for each $n$ we have
a discrete-time filtration $(\Wh^n_i)_{i\geq0}$, an integer $w_n\geq1$
(typically, $w_n\to\infty$), and a sequence $(\de^n_i)_{i\geq1}$ of
random variables.

\begin{lem}\label{LPB1} In the above setting, and if further
each $\de^n_i$ is $\Wh^n_{i+w_n}$-measurable, we have
\bee\label{PB5}
\begin{array}{c}
\E\Big(\sup_{s\leq t}\,\big|\sum_{i=1}^{[s/\De_n]}\de^n_i\big|\Big)~\leq~
K\Big(\frac{a_nt}{\De_n}+2\,\frac{\sqrt{a'_n\,t\,w_n}}{\rdn}\Big)\\
\text{\rm where}~~\quad
a_n=\sup_{i\geq1}\,\E\big(\big|\,\E(\de^n_i\mid\Wh^n_i)\big|\big),\qquad
a'_n=\sup_{i\geq1}\,\E(|\de^n_i|^2).
\end{array}
\eee
\end{lem}

\nib Proof. \rm When $w_n=0$ we have $a_n=\sup_{i\geq1}\,\E(|\de^n_i|)$
and the result is obvious. When $w_n\geq1$ we let
$\de'^n_i=\E(\de^n_i\mid\Wh^n_i)$ and
$\de''^n_i=\de^n_i-\de'^n_i$ and, for $j=1,\cdots,w_n$,
$$A(j)^n_t=\sum_{i=0}^{[(t-j\De_n)/(w_n\De_n)]}\de''^n_{j+iw_n}.$$
The summands above are martingale increments, relative to the
filtration $(\Wh^n_{j+iw_n})_{i\geq0}$, hence by Doob's inequality
$$\E\Big(\sup_{s\leq t}\,|A(j)^n_s|^2\Big)~\leq~\frac{4t\,a'_n}{w_n\De_n}.$$
Observing that $A^n_t=\sdt\de^n_i$
satisfies that
$A_t^n=\sdt\de'^n_i
+\sum_{j=1}^{w_n}A(j)^n_t$, we deduce that
$$\E\Big(\sup_{s\leq t}\,|A^n_s|\Big)~\leq~
\sdt\E\big(|\de'^n_i|\big)+\sum_{j=1}^{w_n}\,
\E\Big(\sup_{s\leq t}\,|A(j)^n_s|\Big),$$
and the result readily follows.\qed
\vsc

\nib 2) \rm Our second auxiliary result mainly compares $R(k_n,\bj)$
defined in (\ref{N4}) with $R(\bj)$ in \eqref{M1}.

\begin{lem}\label{LPB2} If $\bj\in\ja$ we have $|R(k_n;\bj)-R(\bj)|\leq
K/k_n^v$ (for a constant $K$ depending on
$\bj$), hence in particular $R(k_n;\bj)\to R(\bj)$, as $n\to\infty$.
We also have $\Bchi^n_{u_n}\toop0$ for any sequence $u_n$ of integers.
\end{lem}

\nib Proof. \rm Letting $\bj=(j_1,\cdots,j_q)\in\ja$ and $\mu=\mu(\bj)$,
and denoting by $\qa$
the set of all non-empty subsets $Q$ of $\{1,\cdots,q\}$, the complement of
$Q$ being denoted as $Q^c$. (\ref{M1}) and (\ref{N4}) yield
$$R(k_n;\bj)-R(\bj)~=~
\sum_{Q\in\qa}(-1)^{|Q|}\phi_n(Q),\quad
\text{where}~~\phi_n(Q)=\E\Big(\prod_{r\in Q^c}\chi_{j_r}
~\prod_{r\in Q}\Bchi^n_{\mu+(2r-1)k_n}\Big),$$
where $|Q|$ denotes the cardinal of $Q$.
We fix $Q\in\qa$ and let $r_0=\max Q$ and $Q'=Q\backslash\{r_0\}$. Then
$$
\phi_n(Q)=\E\big(\Phi_n(Q)\,\Bchi^n_{\mu+(2r_0-1)k_n}\big),\quad
\text{where}~~\Phi_n(Q)=\prod_{r\in Q^c}\chi_{j_r}
~\prod_{r\in Q'}\Bchi^n_{\mu+(2r-1)k_n}.$$
The variable $\Phi_n(Q)$ is $\g_{\mu+2(r_0-1)k_n}$-measurable, with
$\E(\Phi_n(Q)^2)\leq K$, so by the Cauchy-Schwarz inequality
$$|\phi_n(Q)|~\leq~K\sqrt{\E\big(\big|\E(\Bchi^n_{\mu+(2r_0-1)k_n}
\mid\g_{\mu+(2r_0-2)k_n})\big|^2\big)}.$$
Since $\Bchi^n_{\mu+2(r_0-1)k_n}$
is $\g^{\mu+(2r_0-1)k_n}$-measurable,
centered, and with a second moment bounded in $n$, it follows from
(\ref{P2}) that $|\phi_n(Q)|\leq K/k_n^v$. Summing up over all $Q\in\qa$,
we deduce the first claim.

Finally, we observe that $\E((\Bchi^n_{u_n})^2)$ is independent of $u_n$
and equal to $\frac1{k^2_n}\sum_{0\leq i,j<k_n}r(i-j)$, which in turn
is smaller than $\frac1{k_n}\sum_{m\in\Z}|r(m)|<K/k_n$, and the last
claim follows.\qed
\vsc

\nib 3) \rm For our last auxiliary result we suppose (SNO-2) and consider
$\bj=(j_1,\cdots,j_q)$ and $\bj'=(j'_1,\cdots,j'_{q'})$ in $\ja$, and set
$\mu=\mu(\bj)$, $\mu'=\mu(\bj')$, $\mu''=\mu+\mu'$ and $q''=q+q'$, and
also $\al_n(t)=N_n(t)+1-\mu-2qk_n$ and
$\al'_n(t)=N_n(t)+1-\mu''-(2q''+1)k_n$. The following processes
are the same as $U(\bj)^n$ and $\BU(\bj,\bj')^n$, when there is only
noise and the process $\ga$ is properly ``frozen'':
\bee\label{PS39}
\hskip-2mm\begin{array}{l}
\ua(\bj)^n_t=
\sum\limits_{i=0}^{\al_n(t)}\ga_{i\De_n}^q\,
\prod\limits_{r=1}^q(\chi_{i+j_r}-\Bchi^n_{i+\mu+(2r-1)k_n})\\
\Bua(\bj,\bj')^n_t=
\sum\limits_{i=0}^{\al'_n(t)}\ga_{i\De_n}^{q+q'}\,
\prod\limits_{r=1}^q(\chi_{i+j_r}-\Bchi^n_{i+\mu+(2r-1)k_n})
\prod\limits_{r=1}^{q'}(\chi_{i+\mu+(2q+1)k_n+j'_r}-\Bchi^n_{i+\mu''+(2r+2q)k_n})
\end{array}
\eee

\begin{lem}\label{LPB3} Under (SH) and (SNO-2) we have
\bee\label{PS40}
\left.\begin{array}{l}
\E\Big(\sup_{s\leq t}|U(\bj)^n_s-\ua(\bj)^n_s|\Big)\\
\E\Big(\sup_{s\leq t}|\BU(\bj,\bj')^n_s-\Bua(\bj,\bj')_s^n|\Big)
\end{array}\right\}\leq K_p(t+\sqrt{t})
\big(k_n+\mu''+(k_n+\mu'')^{1/p}\De_n^{1/p-1}\big)
\eee
for any $p>1$, where $K_p$ depends on $p$, and on $\bj,\bj'$ through
$q,q'$ only.
\end{lem}

\nib Proof. \rm Since $U(\bj)^n=\BU(\bj,\emptyset)^n$ and
$\ua(\bj)^n=\Bua(\bj,\emptyset)^n$
(with the convention that an empty
product is equal to $1$, in (\ref{PS39}) for example), only the second
claim needs to be proved.
\vsd

1) The first step is devoted to some estimates. Set for $u,l,w\in\N$:
$$\begin{array}{ll}
\ze(1;u,l)^n_i=X_{(i+u)\De_n}-\BX^n_{i+l},\qquad&\ze(2;u,l)^n_i=
(\ga_{(i+u)\De_n}-\ga_{i\De_n})\chi_{i+u}\\
\ze(3;u,l)^n_i=-\frac1{k_n}\sum_{m=0}^{k_n-1}(\ga_{(i+l+m)\De_n}
-\ga_{i\De_n})\chi_{i+l+m},\qquad&
\ze(4;u,l)^n_i=\ga_{i\De_n}\,\big(\chi_{i+u}-\Bchi^n_{i+l}\big)
\end{array}$$
(note that $\ze(2;u,l)^n_i$ does not depend on $l$ and $\ze(3;u,l)^n_i$
does not depend on $u$).
Upon using the second part of (\ref{P1}) with $V=X$ or with
$V=\ga$, plus the independence of $\f_\infty$ and $\g$ and the fact that
$\chi_i$ has moments of all orders, plus H\"older's inequality, we get for
any $p\geq2$:
\bee\label{PB4}
\begin{array}{ll}
\E\big(|\ze(1;u,l)^n_i|^p\mid\f_{i\De_n}\big)\leq K_p\,\De_n(u+l+k_n),\quad&
\E\big(|\ze(2;u,l)^n_i|^p\mid\f_{i\De_n}\big)\leq K_p\,\De_n\,u\\
\E\big(|\ze(3;u,l)^n_i|^p\mid\f_{i\De_n}\big)\leq K_p\,\De_n(l+k_n),&
\E\big(|\ze(4;u,l)^n_i|^p\mid\f_{i\De_n}\big)\leq K_p.
\end{array}
\eee

One also has the following:
\bee\label{PB6}
j=1,2,3~~\Rightarrow~~\E\Big(\big|\,\E(\ze(j;u,l)^n_i\mid\f_{i\De_n}
\otimes\g)\big|^2\Big)\leq K\De_n^2(u+l+k_n)^2,
\eee
which we prove for $j=3$ only, the cases $j=1,2$ being similar (and even
simpler). Indeed, (\ref{P1}) again and the independence of $\f_\infty$ and
$\g$ yield
$$\big|\,\E(\ze(3;u,l)^n_i\mid\f_{i\De_n}\otimes\g)\big|
\leq\frac{K\De_n(l+k_n)}{k_n}\,\sum_{m=0}^{k_n-1}|\chi_{i+l+m}|.$$
The moments of $\chi$ being finite, one deduces the second part of
(\ref{PB6}).
\vsd

2) By definition, $\BU(\bj,\bj')^n_t-\Bua(\bj,\bj')^n_t=\sum_{i=0}
^{\al'_n(t)}\xi^n_i$, where
$$\begin{array}{lll}
\xi^n_i&=&\prod\limits_{r=1}^q(Y^n_{i+j_r}-\BY^n_{i+\mu+(2r-1)k_n})
\prod\limits_{r=1}^{q'}(Y^n_{i+\mu+(2q+1)k_n+j'_r}
-\BY^n_{i+\mu''+(2r+2q)k_n})\\
&&-\ga_{i\De_n}^{q+q'}\,
\prod\limits_{r=1}^q(\chi_{i+j_r}-\Bchi^n_{i+\mu+(2r-1)k_n})
\prod\limits_{r=1}^{q'}(\chi_{i+\mu+(2q+1)k_n+j'_r}
-\Bchi^n_{i+\mu''+(2r+2q)k_n}).
\end{array}$$

We will rewrite this in a more convenient way. If
$$\begin{array}{llll}
1\leq r\leq q&\Rightarrow&
u_r^n=j_r,&l^n_r=\mu+(2r-1)k_n\\
q<r\leq q''&\Rightarrow&u_r^n=\mu+(2q+1)k_n+j'_{r-q},& l^n_r=\mu''+2rk_n,
\end{array}$$
we have
$$\xi^n_i=\prod_{r=1}^{q''}\big(Y^n_{i+u^n_r}-\BY^n_{i+l^n_r}\big)
-\prod_{r=1}^{q''}\ze(4;u^n_r,l^n_r)^n_i.$$
Since $Y^n_{i+u}-\BY^n_{i+l}=\sum_{j=1}^4\ze(j;u,l)^n_i$,
it follows that, with $\qa$ denoting the set of all partitions
$Q=(Q_1,Q_2,Q_3,Q_4)$
of $\{1,\cdots,q''\}$ such that $Q_1\cup Q_2\cup Q_3\neq\emptyset$,
$$\xi^n_i=\sum_{Q\in\qa}\eta(Q)^n_i,\quad\text{where}~~
\eta(Q)^n_i=\prod_{j=1}^4\eta(Q_j,j)^n_i  ~~\text{and}~~
\eta(Q_j,j)^n_i=\prod_{r\in Q_j}\ze(j;u^n_r,l^n_r).$$
In particular,
\bee\label{PB3}
\E\Big(\sup_{s\leq t}\big|\,\BU(\bj,\bj')^n_s-\Bua(\be,\bj')^n_s\big|\Big)
\leq \sum_{Q\in\qa}N(Q)^n_t,~~\text{where}~~
N(Q)^n_t=\E\Big(\sup_{s\leq t}\Big|\sum_{i=0}^{\al'_n(s)}\eta(Q)^n_i\Big|\Big).
\eee
\vst

3) We now evaluate $N(Q)^n_t$, starting with the case where
$Q$ is such that, among the three sets $Q_1,Q_2,Q_3$, a single one,
say $Q_j$, is a singleton, the other two being empty. We then have
$Q_j=\{r\}$ for some $r\in\{1,\cdots,q''\}$ and $\eta(Q)^n_i=
\ze(j;u^n_r,l^n_r)^n_i\,\eta(Q_4,4)^n_i$. With the variables
$\de^n_i=\eta(Q)^n_i$ and the filtration $\Wh^n_i=
\f_{i\De_n}\otimes\g_{i+\mu''+(2q''+1)k_n}$, and by (\ref{PB4}) and
(\ref{PB6}), H\"older's inequality, and the fact that $\eta(Q_4,4)^n_i$ is
$\Wh^n_i$-measurable, we see that the numbers $a_n$ and $a'_n$
of (\ref{PB5}) satisfy for any $p>1$:
$$a_n\leq K(k_n+\mu'')\De_n,\qquad
a'_n\leq K_{p}\,((k_n+\mu'')\De_n)^{1/p}$$
(with $K,K_p$ depending on $q''$).
One can apply Lemma \ref{LPB1} with $w_n=\mu''+(2q''+1)k_n$ to get
$$N(Q)^n_t\leq
K_p\big((k_n+\mu'')t+\sqrt{t}\,(k_n+\mu'')^{\frac{1+p}{2p}}\,
\De_n^{\frac{1-p}{2p}})$$
(recall that $\al'_n(t)\leq t/\De_n$). Note that by \eqref{M2},
\[
(k_n+\mu'')^{\frac{1+p}{2p}}\De_n^{\frac{1-p}{2p}}
\leq (k_n+\mu'')^{\frac{1}{p}}\De_n^{1/p-1}
\]
for all sufficiently large $n$.

In all other cases of $Q\in\qa$, there are at least two
distinct integers $r$ and $r'$ in $\{1,\cdots,q''\}$ such that $r\in Q_j$ and
$r'\in Q_{j'}$, with $j,j'\leq3$ (we may have $j=j'$). Then
$\eta(Q)^n_i=\ze(j;u^n_r,l^n_r)^n_i\,\ze(j';u^n_{r'},l^n_{r'})^n_i\,
\ze'^n_i$,
and by (\ref{PB4}) and H\"older's inequality we obtain
$\E(|\eta(Q)^n_i|)\leq K_p\,((k_n+\mu'')\De_n)^{1/p}$ for all $p>1$
(with $K_p$ again depending on $q''$). Then in this case
$N(Q)^n_t\leq K_p\,t(k_n+\mu'')^{1/p}\De_n^{1/p-1}$.

These two estimates on $N(Q)^n_t$, according to the case, plus (\ref{PB3}),
imply the second part of (\ref{PS40}).\qed

\subsection{Proof of the Results of Section \ref{sec:M} under (NO-2).}

We begin the proof of the results of Section \ref{sec:M} with
the case of (NO-2), and as written before we can assume the strengthened
versions (SH) and (SNO-2) of our assumptions. We have $N_n(t)=[t/\De_n]$
in this case.

The general idea is to reduce the problem to an application of
Theorem \ref{TPS1}. We fix an arbitrary {\em finite}\ \ subset $\ja_0$ of
$\ja^+$, and if $d=\#\ja_0$ we associate the following
variables $\xi^n$ and $\xi$ and the process $V$, whose components are,
when $\bj=(j_1,\cdots,j_{q(\bj)})$:
\bee\label{PB1}
\xi^{n,\bj}=\prod_{r=1}^{q(\bj)}\big(\chi_{j_r}-\Bchi^n_{\mu(\bj)+
(2r-1)k_n}\big)-R(k_n;\bj),\qquad
\xi^{\bj}=\prod_{r=1}^{q(\bj)}\chi_{j_r}-R(\bj),\qquad
V_t^{\bj}=\ga_t^{q(\bj)}
\eee
By (SNO-2) and Lemma \ref{LPB2}, these variables satisfy (\ref{PS1})
with $w_n=\sup_{j\in\ja_0}(\mu(\bj)+2q(\bj)k_n-1)$ and
$w=\sup_{j\in\ja_0}\,\mu(\bj)$. Note that $w_n^2\De_n\to0$ by (\ref{M2}).
We also write $u'_n=2q(\bj)k_n+\mu(\bj)-1$,
hence $u'_n\De_n\to0$ as well.

With $Z'^n$, $H^n$ and $\ua(\bj)^n$ given respectively by (\ref{M7}),
(\ref{PS5}) and (\ref{PS39}), a simple calculation
shows that
\bee\label{PS60}
\begin{array}{l}
\hskip2cm Z'^{n,\bj}_t=H^{n,\bj}_t+\sum_{l=1}^3 A(l,\bj)^n_t,
\qquad\text{where}\\
A(1,\bj)^n_t=\rdn\,\big(U(\bj)^n_t-\ua(\bj)^n_t\big)\\
A(2,\bj)^n_t=\frac{R(k_n;\bj)}{\rdn}\,\big(\De_n\sum_{i=0}^{N_n(t)-u'_n}
\ga_{i\De_n}^{q(\bj)}-\int_0^t\ga_s^{q(\bj)}\,ds\big)\\
A(3,\bj)^n_t=\frac{R(k_n;\bj)-R(\bj)}{\rdn}\int_0^t\ga_s^{q(\bj)}\,ds.
\end{array}
\eee

Since $U=\ga^{q(\bj)}$ is a semimartingale satisfying (SH), we have
that $\eta^n_i=\rdnn
\int_{i\De_n}^{(i+1)\De_n}(U_s-U_{i\De_n})\,ds$ satisfies
$|\E(\eta^n_i\mid\f_{i\De_n})|\leq K\De_n^{3/2}$ and
$\E((\eta^n_i)^2\mid\f_{i\De_n})|\leq K\De_n^2$, by (\ref{P1}). This, the
boundedness of $\ga_t$, Doob's inequality for {the} discrete-time
martingale $\sum_{i=0}^j(\eta^n_i-\E(\eta^n_i\mid\f_{i\De_n}))$ and
the Cauchy-Schwarz inequality yield
\bee\label{PS66}
\E\Big(\sup_{s\leq t}\,|A(2,\bj)^n_s|\Big)~\leq~K(t+\sqrt{t})\rdn
+Kk_n\rdn
\eee
(the last term in the right being due to $\frac{R(k_n;\bj)}{\rdn}
\int_{(N_n(t)-u'_n)\De_n}^t\ga_s^{q(\bj)}\,ds$).
Finally, we deduce from Lemmas \ref{LPB2} and \ref{LPB3}, and from
the boundedness of $\ga_t$ and the fact that $\E(\|\xi^n\|^2)\leq K$,
that for any $p>1$:
\bee\label{PS62}
\begin{array}{l}
\E\Big(\sup_{s\leq t}\,|A(1,\bj)^n_s|\Big)\leq K_p(t+\sqrt{t})
 \big(k_n\De_n^{1/2}+k_n^{1/p}\De_n^{1/p-1/2}\big)\\
\sup_{s\leq t}\,|A(3,\bj)^n_s|\leq \frac{K t}{k_n^v\,\De_n^{1/2}}.
\end{array}
\eee

Now, we can proceed to the proof of the various results.
\vsc

\nib Proof of (b) of Theorem \ref{TN1}. \rm For (\ref{N2}), it is
enough to check that for all $\bj\in\ja_0$ we have $\rdn\,Z'^{n,\bj}_T\toop0$.
By (\ref{PS60}), this amounts to have $\rdn\,H^{n,\bj}_T\toop0$, which follows
from Theorem \ref{TPS1}, and $\rdn\,A(l,\bj)^n_T\toop0$ for $l=1,2,3$. The
latter is an obvious consequence of (\ref{PS66})--(\ref{PS62}),
plus (\ref{M2}). \qed
\vsc

\nib Proof of Theorem \ref{TM2}. \rm
The covariance $a^{jk}$ of (\ref{PS2}) is denoted by $\Si^{\bj,\bj'}$, and
a simple calculation shows that it is given by (\ref{M5}). Thus the
$\R^{\ja_0}$-valued limit $Z'_T$ in Theorem \ref{TM2} is exactly the limit
$H_T$ in Theorem \ref{TPS1}, as given by (\ref{PS7}). Therefore, in view of
(\ref{PS60}), it is enough to prove that $A(l,\bj)_T^n\toop0$ for $l=1,2,3$
and each $\bj\in\ja_0$. This is an obvious consequence of
(\ref{PS66}) and (\ref{PS62}), upon taking $p=2(1-\theta)$
for the latter, plus the
property (\ref{M12}), which yields in particular $p>1$.\qed
\vsc

\nib Proof of Theorem \ref{TN2}. \rm We have $\bj=(j_1,\cdots,j_q)$ and
$\bj'=(j'_1,\cdots,j'_{q'})$ in $\ja^+$, and we associate the notation
$\mu,\mu',\mu'',q'',\al'_n(t)$ as before (\ref{PS39}).
Instead of (\ref{PB1}) we consider two one-dimensional variables:
$$\xi^{n}=\prod_{r=1}^{q}\big(\chi_{j_r}-\Bchi^n_{\mu+
(2r-1)k_n}\big)-R(k_n;\bj),\qquad
\xi'^{n}=\prod_{r=1}^{q'}\big(\chi_{j_r}-\Bchi^n_{\mu'+
(2r-1)k_n}\big)-R(k_n;\bj').$$
So $\xi^n$ and $\xi'^n$ satisfy (\ref{PS1}), with $w_n=\mu+2qk_n-1$
and $w'_n=\mu'+2q'k_n-1$. Then we associate $\BH^n$ by (\ref{PS20}),
with $V_t=\ga_t^{q''}$ and $\rho_n=k_n+1$ and $u'_n=\mu''+(2q''+1)k_n-1$.

Similar with (\ref{PS60}), we have
$$\begin{array}{l}
\hskip2cm \De_n\BU(\bj,\bj)')^n_t-R(\bj)R(\bj')\int_0^t\ga_s^{q''}\,ds
=\rdn~\BH^n_{t}
+\sum_{l=1}^4A(l)^n_t,\quad\text{where}\\
A(1)^n_t=R(k_n;\bj)\cdot \De_n\,\sum_{i=0}^{[t/\De_n]-u'_n}
\ga_{i\De_n}^{q''}\xi^{n}_i
+ R(k_n;\bj')\cdot \De_n\,\sum_{i=0}^{[t/\De_n]-u'_n}
\ga_{i\De_n}^{q''}\xi'^{n}_{i+w_n+\rho_n},\\
A(2)^n_t=\De_n\,\big(\BU(\bj,\bj')^n_t-\Bua(\bj,\bj')^n_t\big)\\
A(3)^n_t=(R(k_n;\bj)\,R(k_n;\bj'))\,
\Big(\De_n\sum_{i=0}^{[t/\De_n]-u'_n}
\ga_{i\De_n}^{q''}-\int_0^t\ga_s^{q''}\,ds\Big)\\
A(4)^n_t=(R(k_n;\bj)\,R(k_n;\bj')-R(\bj)\,R(\bj'))
\int_0^t\ga_s^{q''}\,ds.
\end{array}$$
Applying Lemma \ref{LPS1} to $H^n$ defined by \eqref{PS5} with
$V_{i\De_n}^j:=\ga_{i\De_n}^{q''}$ and noticing that both $R(k_n;\bj)$ and
$R(k_n;\bj')$ are bounded, we see that $E|A(1)^n_t|\leq K\rdn\sqrt{k_n t}$.
Moreover, using again the boundedness of $R(k_n;\bj)$ and $R(k_n;\bj),$
and if we combine Lemmas \ref{LPS2}, \ref{LPB2} and \ref{LPB3}
(the latter with $p=2$) and also (\ref{PS66}), we
obtain, as soon as $\mu''\leq 2k_n$ and $k_n\De_n\leq1$:
\bee\label{PS63}
\E\Big(\Big|\De_n\BU(\bj,\bj')^n_t-R(\bj)R(\bj')\int_0^t\ga_s^{q''}\,ds
\Big|\Big)
\leq K(1+t+\sqrt{t})\Big(\sqrt{k_n\,\De_n}+\frac1{k_n^v}\Big).
\eee
The right-hand side converges to 0 as $n\to\infty$ under (\ref{M2}),
and the proof is complete.\qed
\vsc

\nib Proof of (b) of Corollary \ref{CN1}. \rm We use the same notation as
in the previous proof. The definitions of
$Z'^{n,\bj}_T$, $\Si^{\bj,\bj'}$ and $\wSi'^{\bj,\bj',n}$, plus the
boundedness of $\ga_t$, yield, with $B_t^n=
\De_n\BU(\bj,\bj')^n_t-R(\bj)R(\bj')\int_0^t\ga_s^{q''}\,ds$
and $r_n=\sum_{m\in\Z:\,|m|>k'_n}|R(\bj\oplus\bj'_{+m})-R(\bj)R(\bj')|$:
\bee\label{PS70}
\Big|\,\wSi'^{\bj,\bj',n}_t-\Si^{\bj,\bj'}\int_0^t\ga_s^{q''}\,ds\Big|
\leq Ktr_n+\De_n^{1/2}\,\Big(|Z_t'^{n,\bj\oplus\bj'}|
+\sum_{m=1}^{k'_n}(|Z_t'^{n,\bj\oplus\bj'_{+m}}|
+|Z_t'^{\bj_{+m}\oplus\bj'}|\Big)+(2k'_n+1)B^n_t.
\eee
Lemma \ref{LPS1} yields $\E(|H^{n,\bj''}_t|)\leq K \sqrt{t}\sqrt{k_n}$,
uniformly in
$\bj''=\bj\oplus\bj'_{+m}$ or $\bj''=\bj_{+m}\oplus\bj'$ when $m\leq k_n$.
Then we apply  (\ref{PS60}), (\ref{PS66}), (\ref{PS62})
with $p=2$, and (\ref{PS63}) to get
$$E\Big(\Big|\,\wSi'^{\bj,\bj',n}_t-
\Si^{\bj,\bj'}\int_0^t\ga_s^{q''}\,ds\Big|\Big)
\leq K(1+t)\Big(r_n+k'_n\big(\sqrt{k_n\De_n}
+1/{k_n^v}\big)\Big).$$
On the one hand, $|R(\bj\oplus\bj'_{+m})
-R(\bj)R(\bj')|\leq K/(|m|-\mu'')^v$ as soon as $|m|>\mu''$, hence
$r_n\to0$. On the other hand,
%%%
%(\ref{N106})
\eqref{N6}
%%%
yields $k'_n(\sqrt{k_n\De_n}+1/k_n^v)\to0$,
and the proof is complete.\qed

\subsection{Proof of the Results of Section \ref{sec:M} under (NO-1).}

Now we turn to the results under (NO-1), and without loss of
generality we can and will assume (SH) and (SNO-1). There is no
process $\ga$ here, but the observation times $T(n,i)$ are (possibly)
random. We also fix the horizon $T>0$.

We consider a finite subset $\ja_0\subset\ja^+$, and we use the
notation (\ref{PB1}), and also the processes $G^n=(G^{n,\bj})_{\bj\in\ja_0}$
defined by (\ref{PS51}), with $u_n=N_n(T)$ and $u'_n=\mu(\bj)+2q(\bj)k_n-1$ and with the variables $\xi^n=(\xi^{n,\bj})$
given by \eqref{PB1} and the corresponding $\xi$ and $w_n$.

\begin{lem}\label{LPC1} For any fixed $T$, the variables $G^n_1$ converge
$\f_\infty$-stably in law to a centered Gaussian $\R^{\ja_0}$-valued
variable $G=(G^{\bj})_{\bj\in\ja_0}$ independent of $\f_\infty$ and
whose covariance matrix is $\Si^{\bj,\bj'}$, as defined by (\ref{M5}).
\end{lem}

In other words, the limit $G$ is the value at time $1$ of the process $B$
of (\ref{PS4}), with the matrix $a=\Si$.
\vsq

\nib Proof. \rm The key point here is the independence between the noise
and the $\si$-field $\f_\infty$. With $B$ as above, we need to
prove that, for any bounded $\f_\infty$-measurable variable $\Phi$ and any
bounded function $f$ on $\R^{\ja}$ which is continuous for the product
topology, we have
\bee\label{PC3}
\E\big(\Phi\,f(G^n_1)\big)~\to~\E(\Phi)\,\E\big(f(B_1)\big).
\eee
In fact, (a) of Theorem \ref{TPS1} can be applied to the (random,
$\f_\infty$-measurable, and going to $\infty$) sequence $u_n=N_n(T)$. We get
that
$$G^{n}_1~\tol~B_1,\quad \text{conditionally on $\f_\infty$,}$$
which in turn yields (\ref{PC3}) in a straightforward manner.\qed
\vsc

Form now on, we basically reproduce the arguments of the previous
subsection. We need to compare the variables $Z^n_T$ of (\ref{M6}) and the
variables $G^n_1$ defined above. If $\al(\bj)_n=N_n(T)+1-\mu(\bj)-2qk_n$
and $\bj=(j_1,\cdots,j_q)$, we observe that
\bee\label{PS651}
\begin{array}{l}
\hskip4cm Z^{n,\bj}_T=G^{n,\bj}_1+A(\bj)_n+A(\bj)'_n,\qquad
\text{where}\\
A(\bj)_n=\sqrt{N_n(T)}~\big(R(k_n;\bj)-R(\bj)\big)\\
A'(\bj)_n=\frac1{\sqrt{N_n(T)}}~\sum_{i=0}^{\al(\bj)_n}\Big(
\prod_{r=1}^{q}(Y^n_{i+j_r}-\BY_{i+\mu(\bj)+(2r-1)k_n})
-\prod_{r=1}^{q}\big(\chi_{i+j_r}-\Bchi^n_{i+\mu(\bj)+(2r-1)k_n}\big)\Big).
\end{array}
\eee
On the one hand, Lemma \ref{LPB2} and the fact that $T/A\leq\De_nN_n(T)
\leq A_T$ (with $A$ and $A_T$ constant) yield
\bee\label{PS65}
|A(\bj)_n|~\leq~K/(k_n^v\,\rdn).
\eee
On the other hand, we have the following lemma, similar to Lemma \ref{LPB3}:

\begin{lem}\label{LPC2} For any $p>1$, and with $K_p$ depending on $p$,
and on $\bj$ through $q(\bj)$ only, we have
$$\E\big(|A'(\bj)_n|\big)\leq K_p\rdn\,
\big(k_n+\mu(\bj)+(k_n+\mu(\bj))^{1/p}\,\De_n^{1/p-1}\big).$$
\end{lem}

\nib Proof. \rm With $\bj=(j_1,\cdots,j_q)$ and $\mu=\mu(\bj)$, we set
$$\begin{array}{l}
\ze(r)^n_i=X_{T(n,i+j_r)}-\BX^n_{T(n,i+\mu+(2r-1)k_n)},\qquad
\ze'(r)^n_i=\chi_{i+j_r}-\Bchi^n_{i+\mu+(2r-1)k_n}.
\end{array}$$
Let $\qa$ be the set of all non-empty subsets $Q$ of $\{1,\cdots,q\}$,
the complement of which being denoted as $Q^c$. We have
$Y^n_{i+j_r}-\BY^n_{i+\mu+(2r-1)k_n}=\ze(r)^n_i+\ze'(r)^n_i$, hence
$$\begin{array}{c}
A'(\bj)_n=\sum_{i=0}^{\al(\bj)_n}\xi^n_i,\quad\text{where}~~
\xi^n_i=\frac1{\sqrt{N_n(T)}}~\sum_{Q\in\qa}\eta(Q)^n_i\,\eta'(Q)^n_i
~~\text{and}\\
\eta(Q)^n_i=\prod_{r\in Q}\ze(r)^n_i,\qquad\eta'(Q)^n_i=
\prod_{r\in Q^c}\ze'(r)^n_i.\end{array}$$

Using $T/A\leq\De_nN_n(T)\leq A_T$ once more, we see that
$$\Big|\,\sum_{i=0}^{\al(\bj)_n}
\eta(Q)^n_i\,\eta'(Q)^n_i\Big|\leq \sup_{t\leq A_T}\,|N(Q)^n_t|,
\quad\mbox{where}~~
N(Q)^n_t=\sum_{i=0}^{[t/\De_n]}\eta(Q)^n_i\,\eta'(Q)^n_i,$$
we see that
\bee\label{PC6}
\Big|\,\frac1{\sqrt{N_n(T)}}~\sum_{i=0}^{\al(\bj)_n}\eta(Q)^n_i\,\eta'(Q)^n_i
\Big|~\leq~\frac{\sqrt{A}}{\sqrt{T}}\sqrt{\De_n}\,\sup_{t\leq A_T}\,|N(Q)^n_t|.
\eee
Note that $\ze'(r)^n_i$ is $\ze(4;j_r,\mu+(2r-1)k_n)^n_i$ of the proof of
Lemma \ref{LPB3}, with $\ga\equiv1$. As for $\ze(r)^n_i$, it is the same as
$\ze(1;j_r,\mu+(2r-1)k_n)^n_i$, except that the $T(n,i)$ are stopping times.
However, since
$T(n,i+m)-T(n,i)\leq mA\De_n$ by (SNO-1), the estimate (\ref{PB4})
and (\ref{PB6}) are still valid here for $\ze(r)^n_i$, with
$\Wh^n_i=\g_{i+j+2qk_n}\bigvee\f_{T(n,i)}$ here. As to $N(Q)^n$, it is
exactly the same here and in (\ref{PB3}). Henceforth, exactly as in this
lemma, we obtain the desired estimate.\qed
\vsc

\nib Proof of (a) of Theorem \ref{TN1}. \rm For (\ref{N1}), it is
enough to check that $Z^{n,\bj}_T/\sqrt{N_n(T)}\toucp0$, whereas
$N_n(T)\asymp 1/\De_n$. This amounts to having $\rdn\,G^{n,\bj}_1\toop0$,
which follows from Lemma \ref{LPC1}, and $\rdn\,A(\bj)_n\toop0$ and
$\rdn\,A'(\bj)_n\toop0$, which follow from (\ref{PS65}) and Lemma
\ref{LPC2}. \qed
\vsc

\nib Proof of Theorem \ref{TM1}. \rm This follows from Lemma \ref{LPC1},
provided $A(\bj)_n\toop0$ and $A'(\bj)_n\toop0$.
Under (\ref{M2}), these two
properties in turn follow from (\ref{PS65}) and Lemma \ref{LPC2}
with $p=2(1-\theta)$, under (\ref{M12}). \qed
\vsc

\nib Proof of (a) of Corollary \ref{CN1}. \rm The definitions of
$Z^{n,\bj}_T$, $\Si^{\bj,\bj'}$ and $\wSi^{\bj,\bj',n}_T$
and the boundedness of $R(\bj)$ and $R(\bj')$ yield, similar
with (\ref{PS70}):
\bee\label{PS71}
\begin{array}{l}
|\wSi^{\bj,\bj',n}_T-\Si^{\bj,\bj'}|\leq r_n+\frac1{\sqrt{N_n(T)}}
\Big(
\,\Big(|Z_T^{n,\bj\oplus\bj'}|+\sum_{m=1}^{k'_n}(|Z_T^{n,\bj\oplus\bj'_{+m}}|
+|Z_T^{\bj_{+m}\oplus\bj'}|\Big)\\
\hskip4cm+\frac{(2k'_n+1)}{\sqrt{N_n(T)}}
\Big(K|Z_T^{n,\bj}|+K|Z_T^{n,\bj'}|
+\frac{|Z_T^{n,\bj}\,Z_T^{n,\bj'}|}{\sqrt{N_n(T)}}\Big)\Big).
\end{array}
\eee
By Lemma \ref{LPS1}, $\E(|G^{n,\bj}_1|)\leq K\sqrt{k_n}$. By (\ref{PS651}),
(\ref{PS65}) and
Lemma \ref{LPC2}, and setting $\de_n=k_n^{1/2}+1/(k_n^v\De_n^{1/2})$, we
conclude $\E(|Z^{n,\bj}_T|)\leq K\de_n$ if $\mu(\bj)\leq2k_n$. Therefore,
since $N_n(T)\geq T/(A\De_n)$, from (\ref{PS71}) and the already proven
fact that $r_n\to0$ (because $k'_n\to\infty$), plus the convergence in
law of $Z^n_T$, we deduce that
$\wSi^{\bj,\bj',n}\toop\Si^{\bj,\bj'}$ as soon as $k'_n\leq k_n$ and
$k'_n\de_n\rdn\to0$. These are implied by (\ref{N6}), and the proof
is complete.\qed

\bigskip
\noindent Jean Jacod: Institut de Math\'ematiques de Jussieu,
4 Place Jussieu, 75 005 Paris, France (CNRS -- UMR 7586, and Universit\'e
Pierre et Marie Curie - P6). jean.jacod@upmc.fr

\medskip
\noindent Yingying Li: Department of Information Systems
 Business Statistics and Operations Management, Hong Kong University of Science and
Technology, Clear Water Bay, Kowloon, Hong Kong. yyli@ust.hk

\medskip
\noindent Xinghua Zheng: Department of Information Systems
 Business Statistics and Operations Management, Hong Kong University of Science and
Technology, Clear Water Bay, Kowloon, Hong Kong. xhzheng@ust.hk

\end{document}